\documentclass[usenatbib]{mn2e}
\usepackage{graphics}
\usepackage{graphicx}
\usepackage{amssymb}
\usepackage{setspace}

\newcommand{\X} {\mathbf{x}}
\newcommand{\V} {\mathbf{v}}

\newcommand{\fxv} {f(\X,\V)}

\newcommand{\funit} {M_{\sun}h^{2}\mbox{kpc}^{-3}\mbox{km}^{-3}\mbox{s}^{3}}
\newcommand{\ten}[2] {#1\times10^{#2}}

\begin{document}

\title{Phase-space structures II: Hierarchical Structure Finder}

\author[M. Maciejewski, S. Colombi, V. Springel, C. Alard, F. R. Bouchet]
{M. Maciejewski$^{1,2}$, S. Colombi$^{1}$, V. Springel$^{2}$, C. Alard$^{1}$, 
F.R. Bouchet$^{1}$\thanks{E-mail: maciejewski.michal@gmail.com; colombi@iap.fr; volker@mpa-garching.mpg.de; alard@iap.fr; bouchet@iap.fr} \\
$^{1}$Institut d'Astrophysique de Paris, CNRS UMR 7095 \& UPMC, 98 bis boulevard Arago, 75014 Paris, France \\
$^{2}$Max-Planck-Institut f\"{u}r Astrophysik, Garching, Karl-Schwarzschild-Stra\ss e 1, 85741 Garching bei M\"{u}nchen, Germany}

\maketitle

\begin{abstract}
A new multi-dimensional Hierarchical Structure Finder (HSF) to study
the phase-space structure of dark matter in $N$-body cosmological
simulations is presented. The algorithm depends mainly on two
parameters, which control the level of connectivity of the detected
structures and their significance compared to Poisson noise.  By
working in 6D phase-space, where contrasts are much more pronounced
than in 3D position space, our HSF algorithm is capable of detecting
subhaloes including their tidal tails, and can recognise other
phase-space structures such as pure streams and candidate caustics.

If an additional unbinding criterion is added, the algorithm can be
used as a self-consistent halo and subhalo finder. As a test, we apply
it to a large halo of the Millennium Simulation, where $19\%$ of the
halo mass are found to belong to bound substructures, which is more
than what is detected with conventional 3D substructure finders, and
an additional $23-36\%$  of the total mass belongs to unbound HSF
structures.  The distribution of identified phase-space density peaks
is clearly bimodal: high peaks are dominated by the bound structures
and show a small spread in their height distribution; low peaks belong
mostly to tidal streams, as expected. However, the projected (3D)
density distribution of the structures shows that some of the streams
can have comparable density to the bound structures in position space.

In order to better understand what HSF provides, we examine the time
evolution of structures, based on the merger tree history. Given the
resolution limit of the Millennium Simulation, bound structures
typically make only up to 6 orbits inside the main halo. The number of
orbits scales approximately linearly with the redshift corresponding
to the moment of merging of the structures with the halo.  At fixed
redshift, the larger the initial mass of the structure which enters
the main halo, the faster it loses mass. The difference in the mass
loss rate between the largest structures and the smallest ones can
reach up to $20\%$. Still, HSF can identify at the present time at
least 80\% of the original content of structures with a redshift of
infall as high as $z \leq 0.3$, which illustrates the significant
power of this tool to perform dynamical analyses in phase-space.
\end{abstract}

\begin{keywords}
methods: data analysis, methods: numerical, galaxies: haloes, 
galaxies: structure, cosmology: dark matter
\end{keywords}

\section{Introduction}
\label{sec:introduction}

When \cite{Zwicky1933} studied galaxy velocities in clusters, he was
the first to notice that there should be about one order of magnitude
more matter in the universe than the observed amount of baryonic
matter to explain the proper motions of galaxies through gravitational
forces. Dark Matter (DM) was introduced to overcome this
problem. Later on, the existence of a dark matter component was
confirmed by the analysis of galaxy rotation curves \citep{Rubin1970}.
Recent studies of gravitational lensing \citep[e.g.][]{lensing} and,
more generally, multiwavelength observations in e.g. the COSMOS
project \citep{COSMOS} provide additional proofs for the existence of
DM. Other constraints on the non-baryonic nature of DM were also set
by the analysis of the Cosmic Microwave Background
\citep[e.g.][]{WMAP5}.

For the last three decades, the DM paradigm has been studied
extensively in the context of cosmological $N$-body simulations. The
comparison of structures formed in such simulations to observed ones
excluded some of the theoretical models, such as Hot Dark Matter
and led to the nowadays commonly accepted $\Lambda$-Cold Dark Matter
($\Lambda$CDM) model. In the $\Lambda$CDM model, dark matter is
collisionless, with a very small velocity dispersion at high redshift;
structures are built in a hierarchical, bottom-up process, where small
structures arise first, seeded from initial fluctuations, and then
merge together to build up larger and larger structures, designated
commonly as {\it haloes}.  Inside the gravitational wells of these
dark matter haloes, baryonic matter forms galaxies \citep{White1978}.

Recently, the efforts to finally identify the physical nature of dark
matter particles, either directly through detecting them in
ground-based dark matter particle detectors or indirectly by observing
their annihilation radiation, have intensified.  At the same time, the
ever increasing resolution of N-body simulations
\citep[e.g.][]{Springel2008} puts new levels of demand on the field of
theoretical study of non-linear halos. The careful analysis of
cosmological structures moves from the study of spherically averaged
three-dimensional density profiles \citep{NFW} to the study of the
full six dimensional phase-space. For example, this concerns the
analyses of the properties of caustics described analytically in
e.g. Bertschinger's secondary infall model \citep{Bertschinger} and
recently reviewed in the context of numerical simulations
\citep{Roya2008,White2008}. Investigations of full phase-space
structures include accurate simulations of two dimensional phase-space
\citep{Alard2005,Colombi2007} and analyses relying on a metric
approach to six dimensional phase-space in $N$-body simulations
\citep{Vogelsberger2008,White2008}. 

A particularly important step in understanding dark matter clustering
lies in an analysis of the bound structures found in $N$-body
simulations. This is at present usually carried out with structure
finders such as {\small SUBFIND} \citep{Springel2001}, {\small
  ADAPTAHOP} \citep{Aubert2004} or {\small PSB}
\citep{Kim2006}. Following this path, we present a new
multi-dimensional Hierarchical Structure Finder which complements all
the above numerical methods with an effective and robust analysis of
phase-space structures in full 6D space.

The paper is organised as follows. First, we review current structure
finders in Section~\ref{sec:structure}.  We then present our new
multi-dimensional Hierarchical Structure Finder (HSF) in
Section~\ref{sec:hsf}. In Section~\ref{sec:test}, we use our algorithm
to detect and analyse phase-space structures of a large halo taken
from the Millennium Simulation. We investigate the space of parameters
on which our HSF algorithm depends and try to find the best choice of
the parameters according to the application under consideration.  We
also introduce the simulation merger tree to follow the evolution of
structures in phase-space.  This allows us to analyse in detail a few
representative cases.  This is followed by a quantitative analysis of
HSF structures in the space and time domain. We also discuss the
bimodal nature of the substructure population, in terms of bound
structures versus tidal tails and tidal streams.  Finally, in
Section~\ref{sec:discussion} we give a summary and present our
conclusions.

\section{Structure finders}
\label{sec:structure}

\begin{figure*}
\includegraphics[width=16cm,height=10cm]{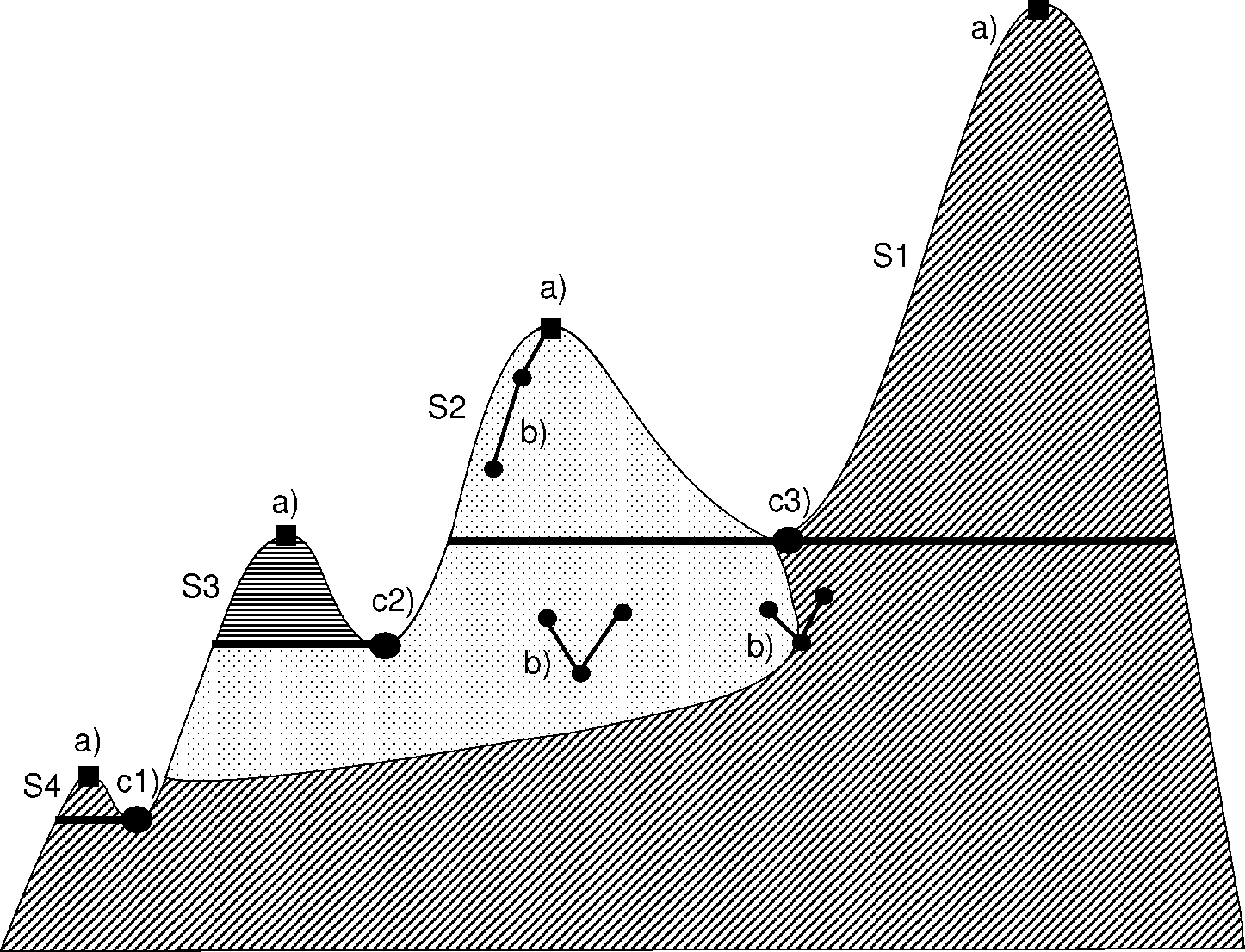}
\caption{Sketch of the hierarchical structure finder. S1, S2, S3 and
  S4 are four different structures found by the HSF algorithm. We
  start to grow structures from local maxima which are marked with
  (a). We grow them down by looking at local neighbours with higher
  density (b), and we connect particles properly to
  structures. Particles on the border of two structures are always
  connected to the larger one. When we find a saddle point (c) we first
  apply our Poisson noise criterion. When the structure is not
  significant (S4), we connect all its particles to the most massive
  partner (c1). If both structures are significant, we apply to them
  the cut or grow criterion, and when one structure is significantly
  less massive than its companion, we grow only the most massive one
  (c2), or if structures have comparable masses, we grow both of them
  (c3).}
\label{hsf}
\end{figure*}

An important step in the analysis of cosmological $N$-body simulations
is to search for virialized dark matter haloes. These are commonly
defined as regions around local density maxima enclosed by a certain
isodensity contour.  The exact definition of such a border changes
from method to method. The simplest and the most popular technique for
finding virialized haloes is the friends-of-friends (FOF) method
\citep{Davis1985}, which links together particles which are separated
by less than a fixed length $b$. Usually $b$ is set to $0.2$ times the
mean inter-particle separation, which corresponds to finding haloes
with overdensity approximately equal to $178$ times the mean
background density $\rho_{\rm mean}$ \citep{Cole1996}.  The mass
function of haloes identified by the FOF method is in good but not
perfect agreement with the predictions of the Press-Schechter
theory. However, the method tends to link together structures across
fine bridges \citep[e.g.][]{Lukic2008} and it is not capable of
detecting substructures inside the virialized haloes themselves.  A
comparable method is the spherical overdensity algorithm
\citep[SO, ][]{Lacey1994} which searches for local density peaks and
then grows around them spheres out to a radius where the enclosed mean
density satisfies a prescribed overdensity criterion. By definition,
the SO method finds only spherical structures. It does not link
structures together with artificial bridges as FOF does, but it may
count mass twice in certain cases.

However, for current high resolution simulations, one needs to find
not only isolated haloes but also their internal substructures. One of
the first methods which made it possible to find such structures is
the hierarchical friends-of-friends scheme \citep{Klypin1999}, in
which a set of different linking parameters, $b$, is used to identify
multiple levels of substructures inside haloes.

To distinguish haloes and their substructures in rich environments,
each detected structure is then usually tested against an additional
binding criterion. This dynamical criterion uses information from
velocity space to guarantee that each structure not only exists but
will survive for a longer period of time.

In the spirit of the SO and FOF methods, the bound density maxima
\citep[BDM,][]{Klypin1999} and DENMAX \citep{Gelb1994} methods were
proposed.  In BDM, particles are grouped in spheres around local
density maxima and are then progressively unbound. In DENMAX,
particles are grouped together when they converge to the same local
density maximum if they are moved along local gradients, calculated on
a rectangular grid. In a final additional step they are attached to
groups identified by the FOF method and then their total binding energy is
checked. This method was generalised in the SKID algorithm
\citep{Governato1997}, in which the local density and its gradient are
calculated directly at the particles positions with the SPH method.  A
similar but simpler method was implemented in the HOP algorithm
\citep{Eisenstein1998}, in which each particle is connected to the one
with the highest density (found by SPH) among its $N_{\rm ngb}$
closest neighbours (with $N_{\rm ngb}$ ranging typically between 
10 and 20). In this way, space is divided into peak-patches that are 
then combined into the final structures.

The HOP method gave rise to new structure finders such as SUBFIND,
ADAPTAHOP, VOBOZ and PSB. The differences between them are sometimes
quite subtle. In SUBFIND \citep{Springel2001}, each particle marks
its two closest neighbours with higher density among the $N_{\rm ngb}$
closest ones. Then particles are sorted by SPH-density in descending
order. Particles without higher density neighbour are marked as the
centres of new structures (local density maxima). Then the structures
grow down in density till they reach border particles, called saddle
points, which have two higher density neighbours belonging to two
different structures.  The smallest structure is marked as a structure
candidate and then both structures are joined together. Structure
candidates are arranged in a hierarchical tree and are successively
unbound, going from bottom of the tree to the top. Particles which are
not bound to a structure are attached to its larger parent structure.

Even though ADAPTAHOP \citep{Aubert2004} constructs the tree of
structures in the opposite way to SUBFIND, from bottom to top, the
main ideas are very similar. First, it grows peak-patches around local
density maxima as in HOP and then finds border particles and saddle
points among them. In addition to SUBFIND, each structure is checked
against Poisson noise to infer its level of significance. In
ADAPTAHOP, contrary to SUBFIND, only particles above saddle points
define structures. VOBOZ \citep{Neyrinck2005} on the other hand uses
Delaunay tessellation to define particle densities and neighbourhood
relations, and also checks the significance level of structures
against a specific Poisson noise criterion. The PSB algorithm
\citep{Kim2006} uses a grid as in DENMAX to find local density maxima
and saddle points, but then constructs a hierarchical structure tree
in the same way as in SUBFIND. In PSB, particles below saddle points
are first attached to all structures which are above them and then are
assigned to individual structures following the process of
unbinding. In addition to the standard unbinding procedure, PSB takes
into account tidal force criteria.

Even though many of the above structure finders use velocity
information for the purpose of a gravitational unbinding procedure,
none of them uses the full six dimensional phase-space
information. However, the advantage of such an approach is that
structures can be defined in a much more natural way in phase-space.
In particular, they have a higher contrast than in position space. In
fact, many structures such as streams and caustics are well defined
only in phase-space. 6D FOF \citep{Diemand2006} is the first
implementation of a structure finder working directly in
phase-space. It is conceptually a simple extension of FOF based on a
six dimensional distance measure, using a fixed global scaling between
position and velocity space. The proposed method finds only local
phase-space density maxima and then grows spheres around them like in
BDM algorithm.

In this paper, we propose a new universal multi-dimensional Hierarchical
Structure Finder (HSF) which is used here to find phase-space
structures in cosmological N-body simulations.  The algorithm
employs, in a higher number of dimensions, a similar approach to
SUBFIND and ADAPTAHOP, but with a new and very effective cut or grow
criterion, controlled by a {\em connectivity parameter} $\alpha$, to
separate accurately structures from each other.

\section{The Hierarchical Structure Finder (HSF)}
\label{sec:hsf}

Our goal is to find the hierarchy of dark matter haloes and subhaloes,
which are defined by locally overdense regions in phase-space. The
main difference between our approach and previous structure finders is
that we focus on all kind of phase-space structures even those which
are not self-bound, such as tidal streams. To enable comparisons, we
however implement, in addition to our base algorithm, also an
unbinding step.  The HSF can be run on a full simulation to detect all
the haloes and the subhaloes population, or on standard groups found
by a FOF algorithm with, e.g., $b=0.2$.

Prior to the identification of structures, our HSF algorithm estimates
the local phase-space density and the local phase-space neighbourhood
of each particle in the sample. Following the proposal of
\cite{Maciejewski2008} for optimum local phase-space density
estimation, we use the SPH method with $N_{\rm sph}$ neighbours found
in the adaptive metric computed by the EnBiD algorithm\footnote{SPH-AM
  in the notation of \cite{Maciejewski2008}}of \cite{Sharma2005}. We
performed a small modification of the EnBiD algorithm to make possible
the output for each particle of the phase-space neighbourhood with
$N_{\rm ngb}$ closest neighbours among the $N_{\rm sph}$ (in the
proper local adaptive metric frame). It is worth mentioning that both
phase-space density and neighbourhood estimations by the EnBiD
algorithm are computationally inexpensive and are almost as fast as
standard three-dimensional SPH estimators.

We define {\it phase-space structures} as the regions grown around
local density maxima by following the local density gradient. To find
such structures, HSF uses a modified version of SUBFIND, which
redistributes particles below saddle points in a new fashion. In the
first step, the HSF algorithm finds locally overdense regions in
phase-space by tracing isodensity contours identified by saddle
points. In addition, we test on each saddle point if structures are
statistically significant when compared to Poisson noise as in
ADAPTAHOP. Particles below phase-space isodensity contours can in
principle be attached to many structures simultaneously but our aim is
to attach each of them to only one structure.  To do that, we use a
simple but robust cut or grow criterion depending on a {\em connectivity
  parameter} $\alpha$, which allows us to reconstruct a multilevel
hierarchy of structures within structures.  In our implementation,
each saddle point defines a connecting bridge between two
structures. Particles below this saddle point are attached to one of
the structures, or are redistributed between both of them according to
the structure masses. We here consider two different cases:
\begin{itemize}
\item the two structures have comparable masses: we attach particles
  below the isodensity contour to both structures in the same manner
  as above the isocontour. Border particles are attached to the most
  massive structure;
\item one structure is significantly less massive than the other one:
  we mark it as a structure and attach all particles below it only to
  the most massive one.
\end{itemize}
While we apply the cut or grow criterion, we create a hierarchical
binary tree of structures by connecting each structure to its more
massive partner if one exists. This way of cutting works like a second
Poisson noise criterion and it allows one to grow only structures
which are significant.

In detail, the HSF algorithm, sketched in Figure~\ref{hsf}, works as
follows:
\begin{enumerate}
\item For each particle, we estimate the local phase-space density
  with SPH-AM and the local adaptive metric environment using $N_{\rm
    ngb}$ neighbours. We usually perform the SPH interpolation with
  $N_{\rm sph}=64$. This value represents a good compromise between
  filtering of Poisson noise and identification of faintest
  significant structures. We find that the final results are rather
  insensitive to the choice of $N_{\rm ngb}$. Our favourite value is
  $N_{\rm ngb}=20$, similar to what is used with HOP, ADAPTAHOP and
  SUBFIND.  Then for each particle, we find the set $A$ of its
  neighbours among the $N_{\rm ngb}$ which have higher density than
  the particle. We sort the set $A$ ascendingly according to local
  neighbourhood distances (closest particles are in the beginning of
  the list). Then we take the two closest elements of $A$ and put 
  them in a second set $B$. This set can be empty or contain one or two
  elements.
\item We sort the particles by decreasing phase-space density and,
  following this ordering, we attach each particle to different
  structures according to the following rules:
\begin{enumerate}
\item The set $B$ is empty. This means that the particle does not have
  any neighbour with higher density: we found a local maximum and we
  mark the particle as the beginning of a new structure.
\item The set $B$ contains one or two particles which belong to the
  same structure: we attach the particle to this structure; or set $B$
  contains two particles which belong to different structures $S_m$
  and $S_n$, and the $S_m$ structure is marked as a more massive
  partner of $S_n$: particle is attached to $S_m$ (this is a border
  particle).
\item The set $B$ contains two particles which belong to different
  structures $S_m$ and $S_n$ and the structures are not on each
  other's more massive partner list: it means that we found a saddle
  point and we perform the marking $S_m>S_n$,  $S_n>S_m$, or
  $S_m\simeq S_n$.

The way this marking is performed in detail can be described as
follows:
\begin{enumerate}
 \item First, we check the level of significance of structures $S_m$
   and $S_n$ when compared to Poisson noise \citep{Aubert2004}. Let
   $\langle S_m\rangle$ and $\langle S_n\rangle$ be the first and the
   second structure's average density and $\rho_{\rm saddle}$ be the
   density of the saddle point connecting them. Each structure is
   significant if
\begin{equation}\label{eq1}
 \langle S\rangle > \rho_{\rm saddle} \left[ 1+\frac{\beta}{\sqrt{N}}\right],
\end{equation}
where $\beta$ is the ``$\beta \sigma$'' level of significance of the
structure (in our tests $\beta$ is set between $0$ and $4$), and $N$
is the number of particles belonging to the structure. If one of the
structures is not significant, then we attach all of its particles to
the second structure. In the case where both structures are not
significant, we attach all the particles to the structure which has
the highest maximum density.
\item If both structures are significant compared to Poisson noise, we
  test them against the cut or grow criterion. Let $|S_m|$ and $|S_n|$
  be the masses of our structures up to this saddle point and $|S_m| >
  |S_n|$, then we mark structure $S_m$ as more massive partner of
  $S_n$. If $|S_m|\alpha>|S_n|$, with $\alpha\in\rbrack0,1\rbrack$,
  then structure $S_n$ is more than $1/\alpha$ times less massive than
  $S_m$ and we attach all the particles below this saddle point to
  $S_m$.

\item If $\frac{|S_m|}{|S_n|} \in
  \rbrack\alpha,\frac{1}{\alpha}\lbrack$, we consider that both
  structures have the same order of mass: we attach the saddle point
  to the most massive structure and all particles below are attached
  according to the rules we set before.
\end{enumerate}
\end{enumerate}

\begin{figure}
\includegraphics[width=8.5cm,height=17cm]{subfind_hsf_bound_unbound.eps}
\caption{Appearance of our Millennium simulation halo (colour ranging
  from green to red, scaling logarithmically with phase-space density)
  and superposed to it, one of its largest substructures found by
  different algorithms (grey pattern).  {\it Left panels:} $x$-$y$
  position space; {\it right panels:} radius $r$ - radial velocity
  $v_r$ phase-space. From top to bottom, the grey pattern corresponds
  to the substructure found respectively by (i) SUBFIND before
  unbinding, (ii) SUBFIND after unbinding, (iii) HSF before unbinding,
  (iv) HSF after unbinding.}
\label{comp_profile}
\end{figure}

\item Finally, a structure containing less than $N_{\rm cut}$
  particles is considered insignificant, and all its particles are
  attached to its more massive partner.  If a structure with less than
  $N_{\rm cut}$ particles does not have a more massive partner, we put
  it on the list of {\it fuzzy} particles.

\item At the end of this process, we obtain a hierarchical tree of
  structures.  Each particle belongs only to one structure or to the
  background (fuzzy list).  In addition, we add to our algorithm a
  final step in which we check each structure against an unbinding
  criterion. Once we have marked its more massive partner for each
  structure, we sort them recursively such that the larger partners
  (parents) are always after the smaller ones (children).  Then we
  unbind structure after structure from children to parents and add
  unbound particles to the larger partner. For each individual
  structure, we calculate the gravitational potential. We set the
  structure centre as the position of the particle with the minimum
  potential and the velocity centre as the mean velocity. We calculate
  the kinetic energy of each particle relative to the mean velocity of
  the structure. All the particles with positive total energy are
  marked and, in that ensemble, $1/4$ of the ones with positive total
  energy are removed. We repeat this process iteratively (starting
  with a new gravitational potential calculation) up to the moment
  when we stay with bound particles only. If the structure has less
  than $N_{\rm cut}$ particles after the unbinding process, then we
  mark it as not bound and attach all its particles to its more
  massive partner or put them on the fuzzy particles list. To speed up
  the calculation of the gravitational potential, we use the tree
  algorithm implemented in {\small GADGET-2}
  \citep{Springel2005GADGET2}.

\end{enumerate}

Most halo finders such as DENMAX, BDM, SKID, SUBFIND, ADAPTAHOP and
VOBOZ use a two step procedure for finding the structures. First, they
assign as many particles as possible to each individual structure in
three dimensional space by tracing local overdensities
(Figure~\ref{comp_profile}, top left panel). When we move to
phase-space diagram (Figure~\ref{comp_profile}, top right panel), we
however immediately observe that there are many particles belonging to
different velocity structures. The unbinding process
(Figure~\ref{comp_profile}, second row of panels) then cleans up all
these spurious velocity structures. In the four bottom panels, one can
observe the results obtained with the six dimensional HSF
algorithm. This method allows us to attach particles to structures in
a more natural way, because it treats both position and velocity space
(Figure~\ref{comp_profile}, third row of panels).  Note that, after
unbinding, the structures detected by the HSF algorithm are more
extended than with standard algorithms working in position space
(Figure~\ref{comp_profile}, bottom panels), an indication that more of
the mass belonging to the substructures is recovered.

\section{Results for a Test halo from the Millennium simulation}
\label{sec:test}


\begin{figure}
\includegraphics[width=8.5cm,height=8.5cm]{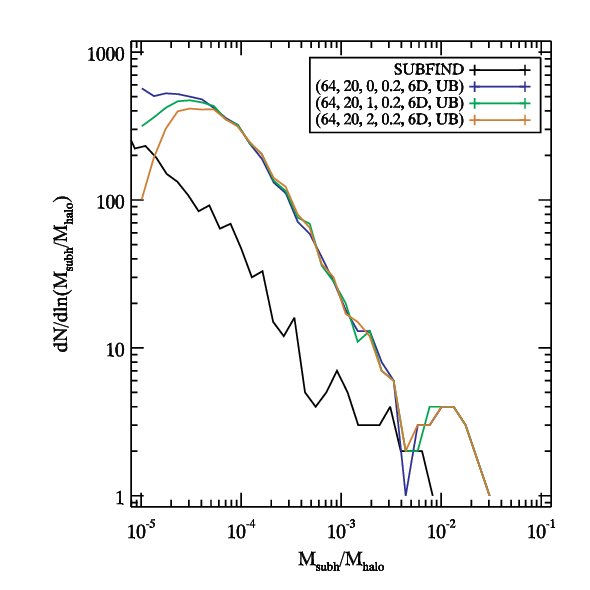}
\caption{Mass distribution of the substructures as a function of the
  ratio between the substructure mass and the mass of the main
  halo. Here, the influence of the choice of the shot noise control
  parameter $\beta$ (Eq.~\ref{eq1}) on the mass profile is tested.}
\label{cutbeta}
\end{figure}

\begin{figure}
\includegraphics[width=8.5cm,height=8.5cm]{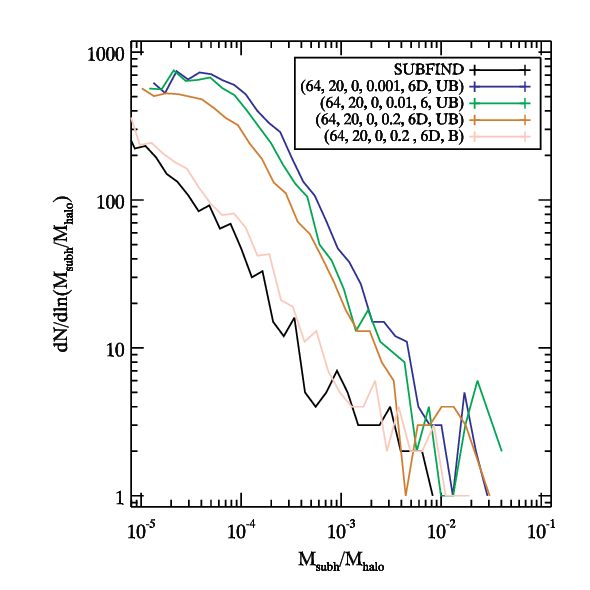}
\caption{Mass distribution of the substructures as a function of the
  ratio between the substructure mass and the mass of the main
  halo. Here, the influence of the connectivity parameter $\alpha$ and
  the importance of the unbinding process are tested.}
\label{cut}
\end{figure}

To test our algorithm we use a large halo extracted from the
Millennium Simulation \citep{Springel2005}.  The main cosmological
parameters of this $\Lambda$CDM-simulation are: $\Omega_m=0.25$,
$h=0.73$, $\Omega_{\Lambda}=0.75$ and $\sigma_8=0.9$
($H_0=100h$~km~s$^{-1}$~Mpc$^{-1}$). The simulation volume is a
periodic box of size $500\,h^{-1}{\rm Mpc}$ and individual particles
have a mass $8.6\times10^8\,h^{-1}{\rm M}_{\sun}$.  In our analysis we
take the second largest FOF halo at redshift $z=0$, which has
$3.83\times10^6$ particles.

This section is organised as follows. In Section~\ref{sec:choice}, we
discuss the influence of the main parameters in our algorithm on the
results. In Section~\ref{sec:phasespace} we use the merger tree
history to follow both qualitatively and quantitatively the evolution
of structures. In particular, the structures identified by HSF are
cross-correlated with their counterpart prior to merging with the main
halo.  Finally Section~\ref{sec:bound} studies the properties of the
substructure population obtained with HSF and its bimodality in terms
of bound structures versus unbound tidal tails and tidal streams.

\subsection{Choice of the main parameters in the algorithm} 
\label{sec:choice}

In the following, we check the influence of the different parameters
on the structures found by our HSF algorithm. A basic parameter setup
is given by $N_{\rm sph}=64$, $N_{\rm ngb}=20$, $\beta=0$,
$\alpha=0.2$, $N_{\rm cut}=20$.  We adopt the notation $(N_{\rm
  sph},N_{\rm ngb}, \beta, \alpha, {\rm dimension}, {\rm
  (B)ound/(UN)bound)}$ to label each set of parameters. When the
dimension is set to 3D, we mean the three dimensional position space,
whereas 6D means six dimensional phase-space. Unless mentioned
otherwise, we use the HSF algorithm without additional unbinding step.
SUBFIND is in fact one of the versions of our algorithm, characterised
by the following parameter setup ($64,20,0, \mbox{always cut smaller
  partner}$, 3D, B).  In our analysis, we shall call all the particles
in a FOF group {\it a halo}, the largest structure of the FOF group
{\it a main halo}, and all other structures {\it substructures}.

Figures~\ref{cutbeta} and \ref{cut} present the number of
substructures per logarithmic mass bin scaled to the main halo
mass. While testing the different parameter setups, we find that
$\alpha$ and $\beta$ are the most influential. Figure~\ref{cutbeta}
shows that parameter $\beta$ is important for the smallest structures:
visual inspection suggests that a higher $\beta$ helps to preserve
small scale connectivity, e.g. between tidal tails and the bound
component of the substructures. The connectivity parameter $\alpha$
has a similar effect, but on the whole mass range instead of small
structures only, as illustrated by Figure~\ref{cut}. Because of the
partial degeneracy between $\alpha$ and $\beta$, we prefer to use
$\beta=0$ in most of our analyses. The choice of $\alpha$ indeed
influences connectivity as follows: when $\alpha=0.2$, the main halo
always wins the cut or grow criterion and all structures are cut by
it; when $\alpha=0.02$, the largest substructures can grow inside the
main halo; when $\alpha=0.01$, all small substructures grow more
aggressively and the halo is divided into more small parts. In brief,
focusing on bound structures calls for a value of $\alpha$ of the
order of $0.2$, while if one is interested in all substructures
including tidal streams, it is better to set
$\alpha\simeq0.01-0.02$. In the later case, tuning up $\beta$ can help
to control the small scale connectivity.

\begin{figure}
\includegraphics[width=8.5cm,height=8.5cm]{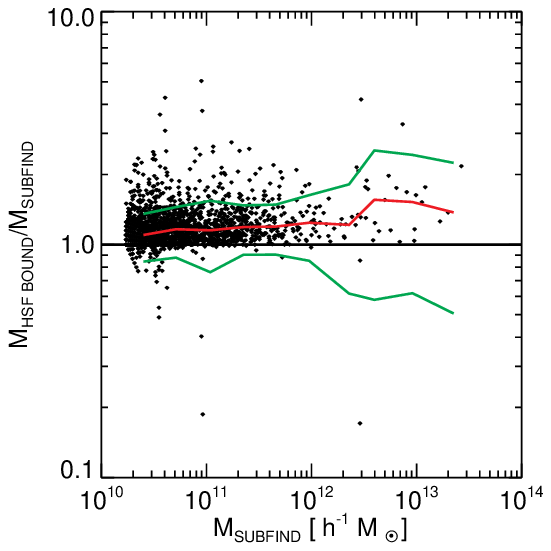}
\caption{Ratio of mass of each HSF bound structure divided by the mass
  of its SUBFIND counterpart as a function of SUBFIND structure
  mass. The central curve corresponds to the median value of the ratio
  calculated over 10 logarithmic bins along the $x$-axis, taking into
  account only bins containing 2 points or more. The two additional
  green curves on each side show $1\,\sigma$ errors estimated from the
  variance of points in each bin.}
\label{comp}
\end{figure}

Using our base parameter setup, we now compare HSF bound structures
with those given by SUBFIND.  The HSF algorithm works in six
dimensional phase-space, where the distribution of particles shows
much more contrast than in position space alone. Because of that, HSF
can better trace contours of individual substructures and attach more
particles to them. Even after the unbinding step, HSF therefore
attaches more particles to the substructures than SUBFIND. This is
illustrated by Figure~\ref{comp}, where the ratio between the mass of
HSF bound structures and the mass of their SUBFIND counterparts is
plotted: HSF attaches on average $\sim 1.1$ more mass to small
structures than SUBFIND and up to twice more to the largest ones.

\begin{figure*}
\includegraphics[width=18cm,height=18cm]{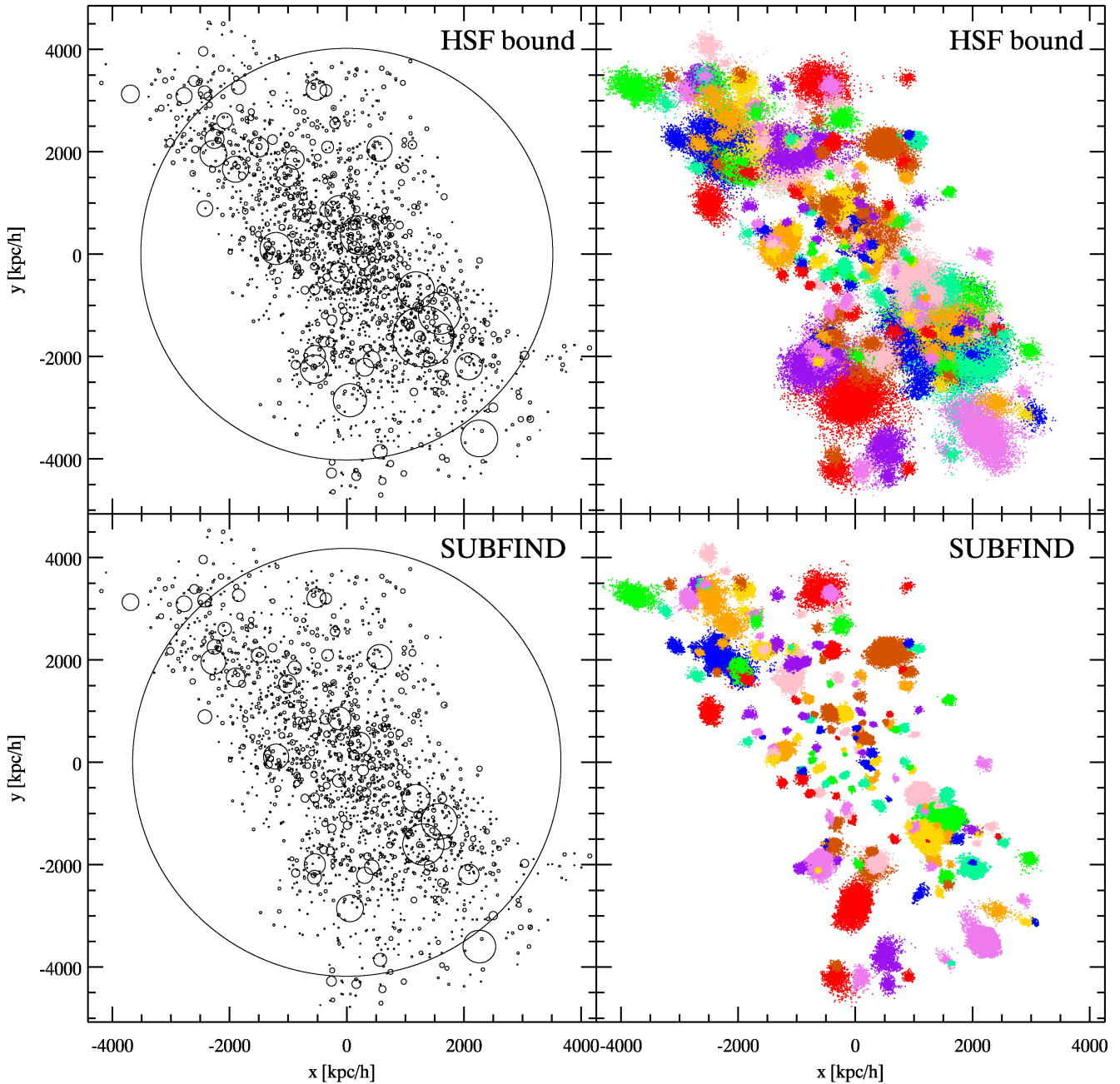}
\caption{{\em Left side panels}: Spatial distribution of HSF bound
  structures and SUBFIND structures, with at least $20$ particles. The
  area of each circle is proportional to the structure mass.  {\em
    Right side panels}: First $200$ most massive bound substructures
  identified by HSF and their SUBFIND counterparts. Particles
  belonging to the same substructure share the same colour.}
\label{subfind_mass}
\end{figure*}

The left panels of Figure~\ref{subfind_mass} compare bound structures
found by both methods in position space.  The area of each circle is
proportional to the structure mass. 

With the parameters set up chosen here, HSF finds around $10\%$ more 
structures, mostly small ones, in the outskirts of the main halo and 
clearly attaches more mass than SUBFIND to most of them. Nevertheless, 
the spatial distributions of HSF and SUBFIND substructures are nearly 
the same, as expected. To complete this visual inspection, the right panels of
Figure~\ref{subfind_mass} compare the 200 largest bound structures
found by both methods. Most of the HSF structures are matched by
SUBFIND, except that they are more extended. As mentioned before, many
of these structures have $1.1-2$ times more mass in HSF than in
SUBFIND. Interestingly, this confirms the mass excess found around
SUBFIND substructures in a comparison of simulation with gravitational
lensing observations by \cite{Natarajan2007}.

\begin{figure}
\resizebox{8.5cm}{!}{\includegraphics{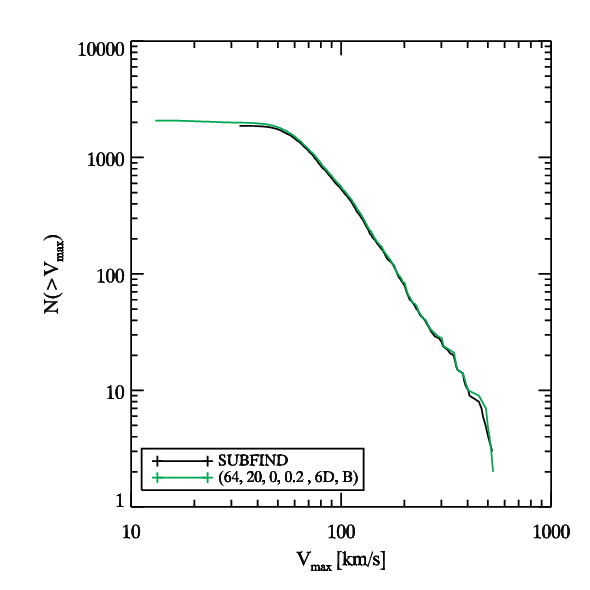}}
\caption{Cumulative count of substructures as a function of maximum
  circular velocity. We here compare results of HSF for bound
  structures with substructures identified by SUBFIND in the same
  halo.}
\label{velmax}
\end{figure}

If bound substructures are counted as a function of maximum circular
velocity instead of mass, a much closer agreement is found,
however. This is seen in Figure~\ref{velmax}, where the cumulative
velocity functions of bound substructures for HSF and SUBFIND are
compared. HSF tends to find a few more small substructures, but both
algorithms essentially identify the same set of more massive
structures, confirming the results above.

\subsection{Phase-space structures and merger tree}
\label{sec:phasespace}

In the following sections we describe a method to follow back in time
the structures detected by our HSF, paying particular attention to the
definition of what we use for initial halo. Then we study in detail a
set of specific but representative cases.  The goal of this analysis
is to physically understand the nature of the structures found by our
algorithm, before and after unbinding. In particular, we aim to
separate clearly tidal streams from compact bounded subhaloes. With
the additional time information, we can also associate tidal streams
to objects in the stage they were prior to merging with the main
halo. We can also study quantitatively how in general phase-space
structures evolve in time.

\subsubsection{Evolution with time and the merger tree}
\label{sec:evolution}

To study in detail the nature of phase-space structures found by the
HSF algorithm we use the merger tree history\footnote{Each branch of
  this tree corresponds to the evolution in time of a dark matter halo
  as a stand alone structure, while each node of it corresponds to the
  event of merging between 2 halos or more. Note that, due to the
  collisionless nature of dark matter, the halos can pass through each
  other and separate again: in practice, the structure of such a tree
  can be non-trivial.} to follow their evolution backwards in
time. Then we count how many particles are shared between each
structure prior to merging with the main halo and its counterpart
detected by HSF at $z=0$. This process uses pieces of information
which are already available for the processed Millennium Simulation
\citep{Springel2005b} and it is divided into three steps: (i)
crosscorrelating the HSF structure catalogue with the SUBFIND one,
(ii) following the evolution of SUBFIND structures using the already
implemented merger tree history and (iii) using each particle's
universal index\footnote{The universal index of a particle is just a
  number associated to each single particle in order to identify it
  unambiguously, which is useful for analyses of Lagrangian nature
  such as performed in this work.} to follow structures at different
output times. We now explain each of these steps in turn.

\begin{enumerate}

 \item
cross-correlation between HSF and SUBFIND: Information about
structures in the Millennium Simulation is organised in terms of two
levels: first, particles are attached to different FOF groups (found
with $b=0.2$). Then, in each FOF group, they are separated into the
main halo, the substructures found by SUBFIND, and unbound `fuzzy'
particles if present.  Running the HSF algorithm with the base
parameter setup $(64, 20, 0, 0.2,$ 6D UB/B$)$ provides a phase-space
structure list. Then, for each member of that list, the SUBFIND
substructure sharing the largest possible number of particles with it
is identified. If the HSF structure shares less than $20$ particles
with any SUBFIND substructure, it is put into an unmatched list. In the
opposite case, we call this SUBFIND substructure a {\it seed} of the
HSF structure.


\item

Following SUBFIND structures back in time: Information about the time
evolution of structures is stored in the Millennium Simulation in a
merger tree \citep[more details in][]{Springel2005b}. We use the tree
information which gives for each halo or substructure its most massive
progenitor, if there is any. Once the list of seed SUBFIND
substructures is obtained, each of them is traced back in time by
following its most massive progenitor recursively up to the moment
when this past structure was the main halo of a FOF group. This is the
last occurrence of the structure as being distinguishable as an
isolated halo. We store the redshift of this event and all particles
belonging to the main halo found in this way are denoted as the {\it
  initial halo}. There is a small number of substructures which do not
have a proper progenitor, they are dropped from the analysis.

\item
Using each particle's universal index to follow structures at
different times: In the last part of the procedure, we link together
the information gathered in the previous two steps. For each HSF
structure identified at the present time, we find its initial halo
and, with the help of universal indices, we identify {\it shared}
particles, i.e.~particles belonging both to the initial and final
structures.  We carry out exactly the same analysis for HSF bound
structures and for SUBFIND itself.

\end{enumerate}

In addition, during this process, we gather for each substructure
information about the position of its centre and its velocity at
various times (we use the SUBFIND definition for the structure
centre), and similarly for the position and velocity of the centre of
the main halo. With these pieces of information at hand, we can define
an orbital count by determining the number of times a substructure's
radial velocity with respect to the centre of the halo changes sign,
which each time signals completion of what we call an orbit.
This
definition requires that there are enough snapshots to catch orbit
details. This is the case for most substructures, probably all,
although this statement is not easy to check rigorously.

\subsubsection{Definition of the initial halo}
\label{sec:definition}

\begin{figure*}
\includegraphics[width=17.0cm,height=21cm]{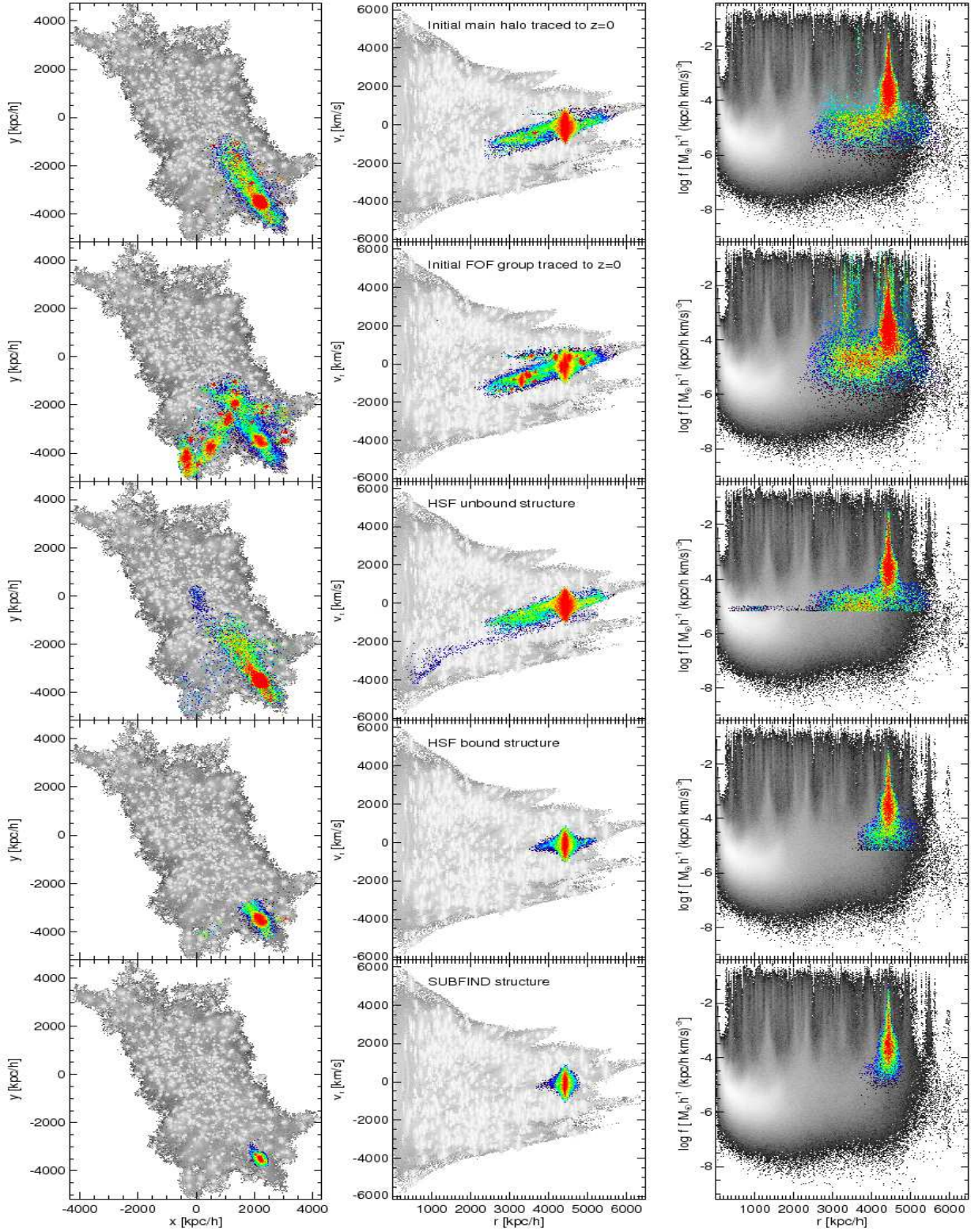} 
\caption{Follow up of a particular substructure (in colour) of our
  Millennium test halo (in grey).  {\em From left to right:} $x$--$y$
  position space, radius $r$--radial velocity $v_r$, radius
  $r$-phase-space density $f$.  In the two left columns of panels, the
  colour traces the logarithm of the phase-space density (from dark
  grey to light grey or from dark blue to red).  In the right column
  of panels, the colour just traces the projected particle density.
  {\em From top to bottom:} Initial main halo traced to $z=0$, initial
  FOF group traced to $z=0$, HSF unbound structure, HSF bound
  structure, and SUBFIND structure.  }
\label{FOF}
\end{figure*}

In our analysis of the time evolution of the structures, we adopt the
main haloes of the FOF groups found by SUBFIND as `initial haloes'.
Another possibility is to chose for each initial halo all the
particles belonging to the FOF group.  In the first case, the analysis
is simplified by the fact that we look only for the evolution of one
isolated component.  However this component represents only one part
of the halo. In the second case, a FOF group can sometimes have a few
main halo candidates joined by small artificial bridges \citep[up to
  $20\%$ of FOF groups show such a feature, e.g.][]{Kim2006} but can
in fact be tidally disrupted such that its components get away from
each other.

To demonstrate the effects described above, we choose one particular
structure in our test halo. The top panel of Figure~\ref{FOF} shows
all the particles belonging to the initial halo traced to redshift
$z=0$, while the second row of panels corresponds to the full traced
initial FOF group. This structure goes around the main halo one time
(its orbit is shown in the third row of panels of
Figure~\ref{halo_profile}).  The initial FOF group is tidally
disrupted during this process and its various components are clearly
separated from each other. SUBFIND recognises the central part of the
bound object (bottom panels of Figure~\ref{FOF}). The HSF bound
structure contains more particles (fourth row of panels in
Figure~\ref{FOF}). These particles belong to tidal tails, but are in
fact still gravitationally linked to the structure. The HSF structure
(prior to unbinding) contains $55 \%$ of the particles of the initial
halo, and reproduces perfectly its shape (on third row of
Figure~\ref{FOF}).

\subsubsection{Qualitative analysis of structure evolution}
\label{sec:qualitative}
\begin{figure*}
\includegraphics[width=18cm,height=20cm]{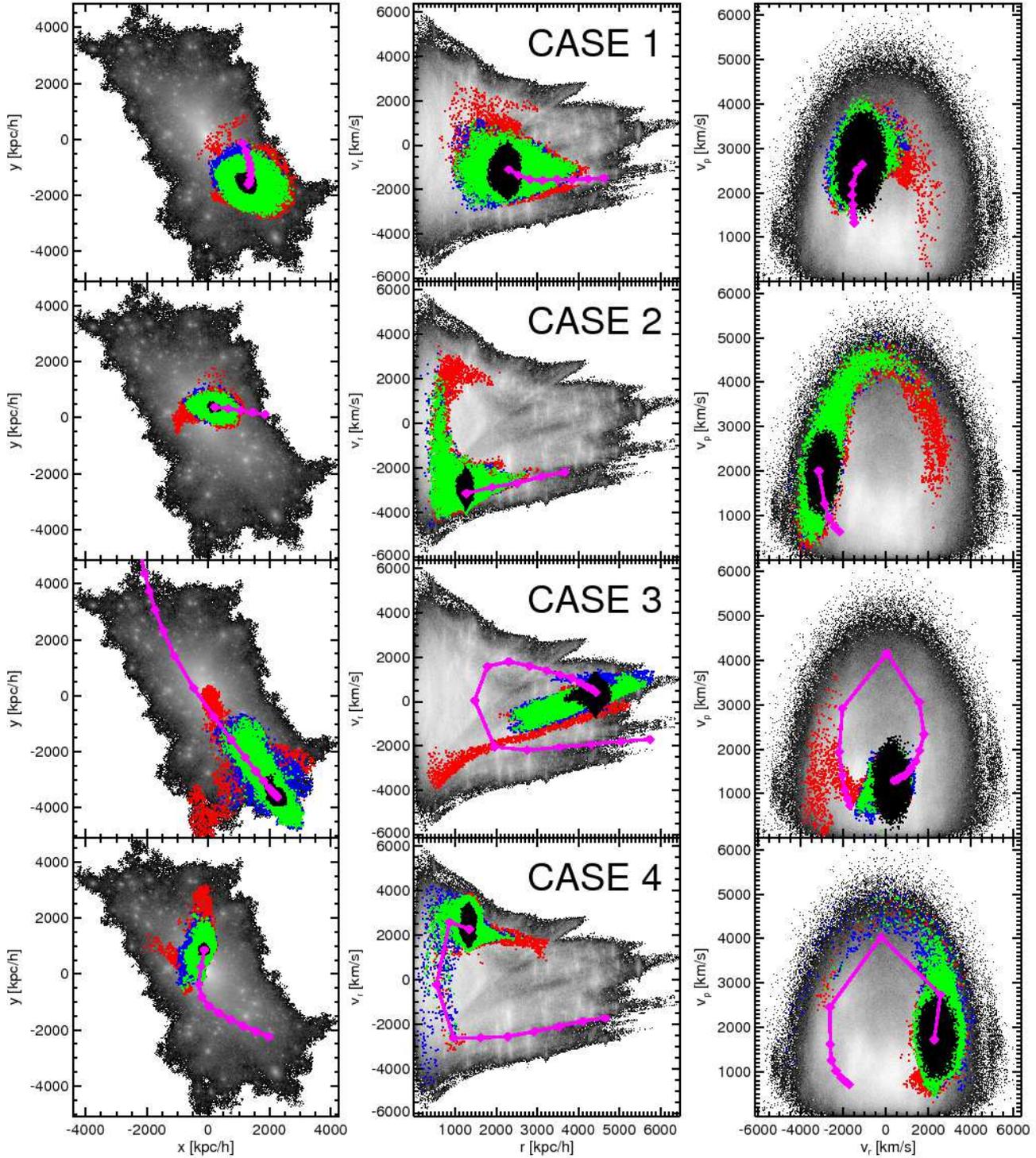}
\caption{Properties of some chosen structures. From left to right:
  $x$--$y$ position space diagram, radius $r$--radial velocity $v_r$
  diagram, radial velocity $v_r$--tangential velocity $v_t$ diagram.
  Further description in the text (section \ref{sec:qualitative}).}
\label{halo_profile}
\end{figure*}
\addtocounter{figure}{-1}
\begin{figure*}
\includegraphics[width=18cm,height=20cm]{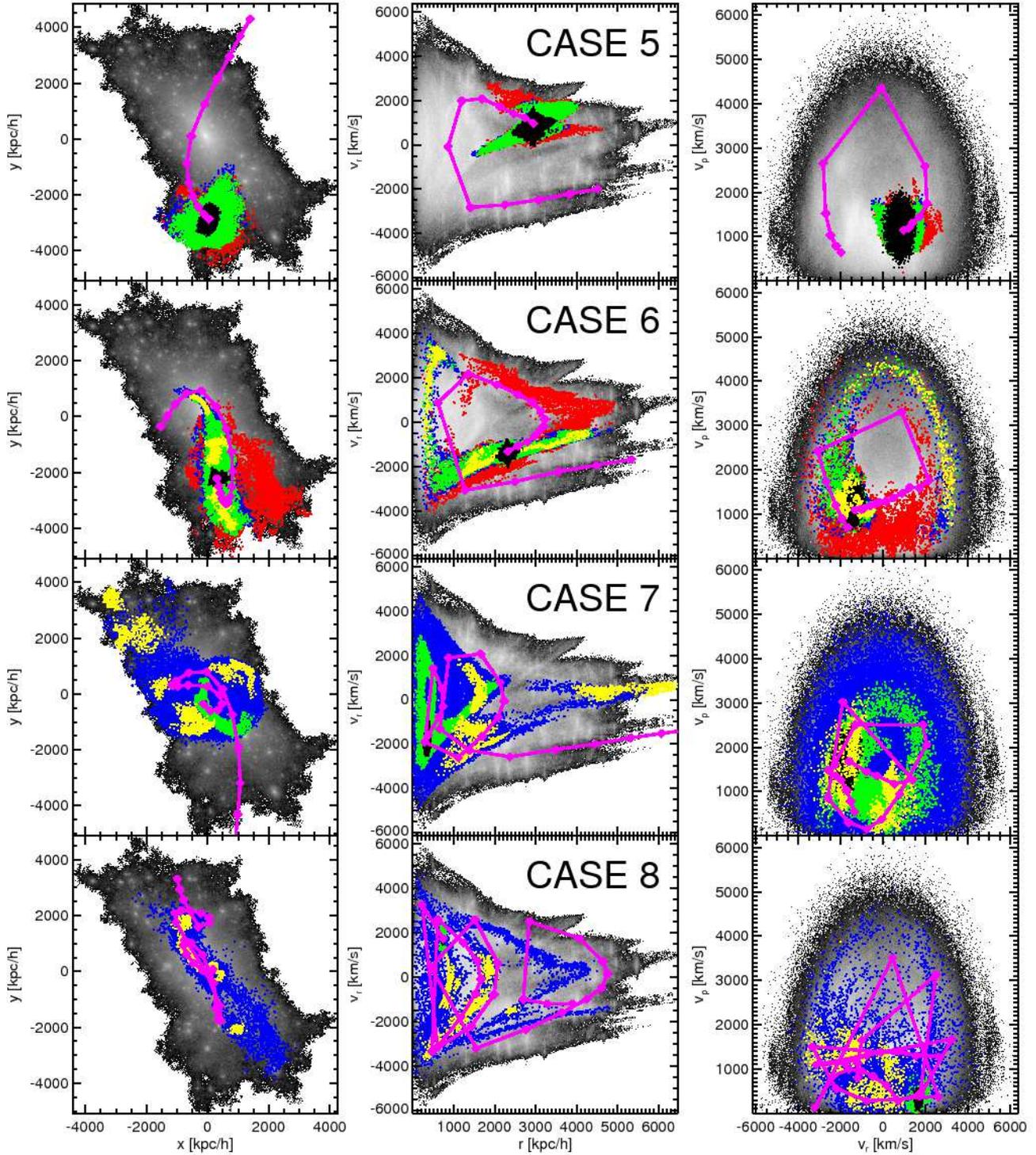}
\caption{Continued.}
\end{figure*}
To better understand all the processes at play during structure
evolution, we study in greater detail eight different cases displayed
in Figure~\ref{halo_profile}.  The colours in the figure are coded as
follows:
\begin{itemize}
 \item Green and red particles belong to one structure found by the
   HSF algorithm (without unbinding): green particles belong to the
   initial halo, while the red particles do not belong to it.
 \item Black particles belong to the SUBFIND seed of the HSF
   structure.
 \item Blue particles belong to the initial halo, but do not belong to
   the HSF structure.
 \item Yellow particles belong to the initial halo and belong to any
   HSF structure, besides the one we take for the current analysis. We
   mark particles in yellow only for structures in the last three rows
   of panels of Figure~\ref{halo_profile}.
\end{itemize}
The particles are plotted in the following order: blue, red, green,
black, and finally yellow.  Various structures parameters are listed
in Table~\ref{strpr}, for each of the 8 cases considered here. The
pink curve shows the orbit of the object inside the main halo. We now
discuss in detail each of these 8 cases.

\begin{table*}
\caption{Properties of eight chosen structures. Column description: 
{\em nr.:} structure number, in the same order as listed in the text and in Figure~\ref{halo_profile}; 
{\em orbits:} number of orbits inside the main halo;
{\em HSF par.:} number of particles belonging to the HSF structure; 
{\em Initial FOF group par.:} number of particles belonging to the initial FOF group; 
{\em Initial halo par.:} number of particles belonging to the initial main halo;
{\em SUBFIND:} fraction of the particles of the initial halo that are still identified 
in SUBFIND (in HSF bound, HSF, any identified HSF structures, respectively for the 
next columns).}
\begin{tabular}{|c|c|c|c|c|c|c|c|c|c|}\hline
nr. & orbits & z & HSF par. & Initial FOF group par. & Initial halo
par. &SUBFIND &HSF bound & HSF & any HSF \\ \hline \hline
$1$ & $0$ &$0.17$ & $78791$ & $109835$ & $76567$ & $37\%$ & $81\%$ &$84\%$ & $94\%$\\    
$2$ & $0$ &$0.09$ & $37417$ & $39511$  & $32383$ & $26\%$ & $84\%$ &$97\%$ & $99\%$\\    
$3$ & $1$ &$0.56$ & $36248$ & $85225$  & $53186$ & $35\%$ & $44\%$ &$55\%$ & $57\%$\\  
$4$ & $1$ &$0.24$ & $35520$ & $43202$  & $35463$ & $26\%$ & $43\%$ &$88\%$ & $91\%$\\  
$5$ & $1$ &$0.32$ & $33647$ & $30432$  & $28604$ & $44\%$ & $60\%$ &$90\%$ & $92\%$\\  
$6$ & $1$ &$0.62$ & $17971$ & $17053$  & $14364$ & $32\%$ & $38\%$ &$65\%$ & $72\%$\\  
$7$ & $2$ &$0.83$ & $14477$ & $140799$ & $119621$& $2\%$  & $5\%$  &$11\%$ & $12\%$\\    
$8$ & $4$ &$1.50$ & $854$   & $10696$  & $9896$  & $3\%$  & $6\%$  &$8\%$  & $25\%$\\ \hline   

\end{tabular}
\label{strpr}
\end{table*}

\begin{enumerate}
 \item The first case corresponds to the largest structure found by
   HSF.  It entered the main halo recently, at redshift $z=0.17$.  HSF
   identifies a large fraction of $84\%$ of the initial structure, and
   unbinding worsens that number by only about $3\%$.  On the other
   hand, SUBFIND identifies only $37\%$ of the content of this rather
   massive structure.
 \item The second structure is at a moment just before crossing for
   the first time the main halo centre and starts to be significantly
   tidally disrupted. The HSF structure still contains $97\%$ of
   particles of the initial main halo, while the SUBFIND bound
   structure accounts for only $26\%$. Indeed, HSF manages to attach
   to the structure unbound particles which already crossed the main
   halo centre and contribute to a tidal tail.
 \item The third structure entered the halo at redshift $z=0.56$ and
   made an orbit inside it. This is the reason why we identify
   only $55\%$ of the initial structure, but still more than SUBFIND
   ($35\%$).  HSF attaches some additional particles to the structure,
   i.e.~particles that do not belong to the initial main halo, but in
   fact we found that a large fraction of them belong to the initial
   FOF group.
\item The fourth case corresponds to a structure that just passed the
  main halo centre.  $88\%$ of the initial structure is still
  identified. HSF is capable of connecting to the structure (unbound)
  particles which were strongly affected by tidal stripping but still
  belong to the structure. In this case for example, it joins some
  particles having high negative velocity.
 \item The fifth case corresponds to the example of a structure that
   has gone around the main halo centre. HSF identifies $90\%$ of the
   initial structure and SUBFIND only $44\%$. We can clearly see here
   the typical ``z'' shape of the structure in $(r,v_r)$ space, which
   correspond to a scaled down version of the phase-space diagram of
   the halo (that we see only in half here).
 \item The sixth row of panels corresponds to a rare occurrence when
   HSF partly fails. The HSF structure contains $65\%$ of particles of
   the initial one. It is falling inside the main halo centre and some
   of the particles already crossed the centre. Parts of the initial
   structure are identified as other HSF objects: the initial halo
   shares $72\%$ of its particles with all detected HSF substructures.
   There is however a large fraction of particles associated with the
   HSF structure (in red) that should just belong to the
   background. Indeed, we checked that most of them cannot even be
   associated with the initial FOF group.
 \item The seventh row of panels shows the typical case of a massive
   structure which, after only two orbits (so it passed nearby the
   halo centre only twice), already dissolved in the main halo, because
   of massive tidal disruption. Even though the HSF structure contains
   only $11\%$ of the particles of the initial halo, we find that
   other HSF substructures match some parts of the initial halo:
   $90\%$ of particles inside such substructures come from the initial
   halo, although some of them belong to other members of the initial
   FOF group. In other words, it means that our algorithm is capable
   of finding remnants of tidal tails. All blue particles can not be
   distinguished from the main halo anymore and correspond to the part
   of the structure which has been completely diluted.
 \item In the last case, we take a structure which merged with the main halo at 
       high redshift $z=1.5$, and already made four orbits inside. As expected,
       this structure has experienced strong tidal striping: the HSF structure
       contains only $8\%$ of the initial structure. Most of it 
       indeed already dissolved inside the main halo. Still, some remnants were 
       identified as disjoint HSF components and represent $25\%$ of 
       the initial structure.
\end{enumerate}

To conclude this section, objects which recently entered the main halo
and typically made up to one orbit inside it are in most cases
fully recovered by HSF. When the structures make more orbits, they are
more tidally disrupted, especially when they come close to the halo
centre, so the fraction of particles identified decreases. HSF still
finds in most cases remnants of strongly disrupted objects as
individual tidal streams detached from the bound component (if this
later still exists).

\subsubsection{Quantitative analysis of structure evolution}
\label{sec:quantative}

\begin{figure*}
\includegraphics[width=17cm,height=17cm]{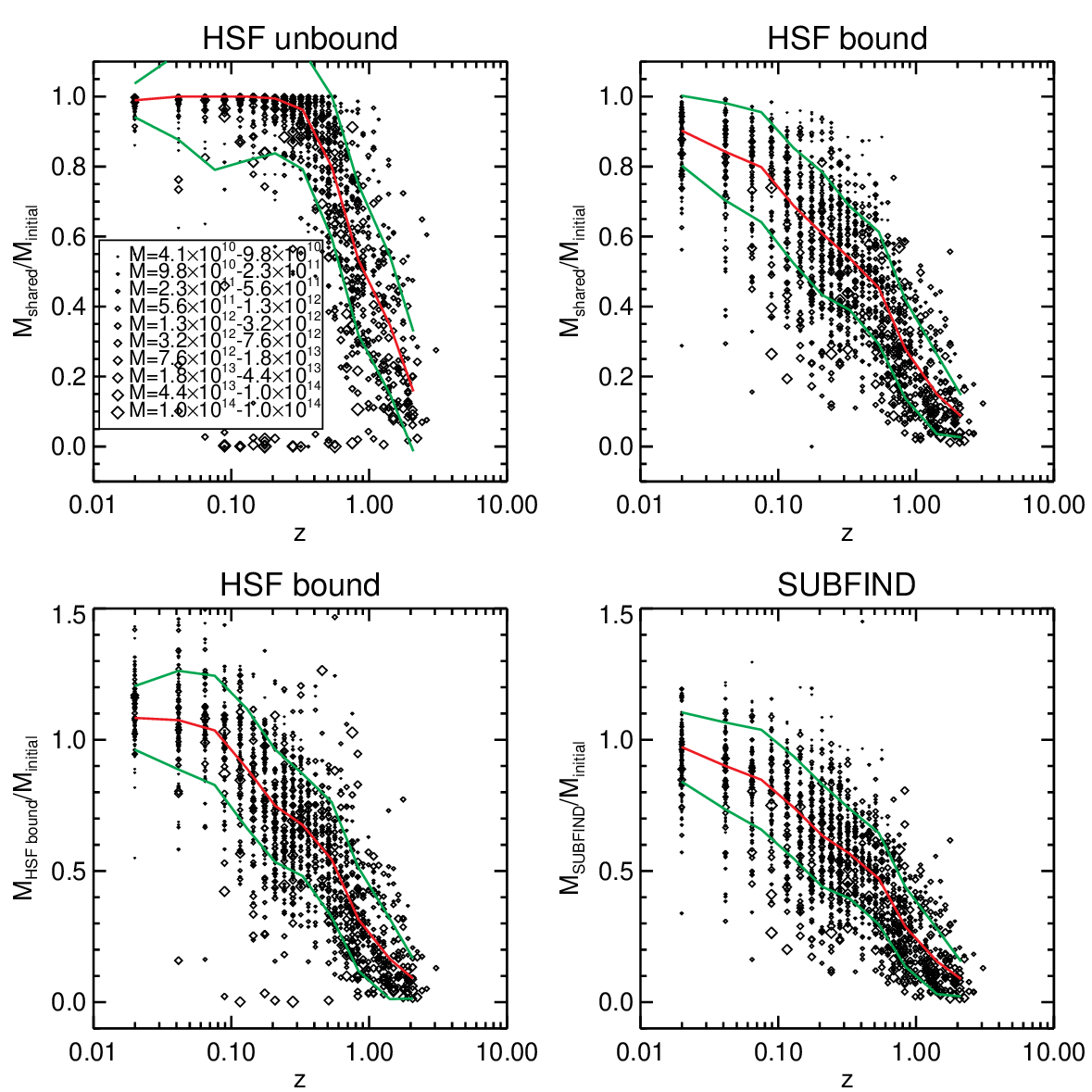}
\caption{Mass loss of structures as a function of the redshift $z$ of
  merging with the main halo. {\em Top-left panel:} the mass loss is
  presented as the ratio $M_{\rm shared}/M_{\rm initial}$, where
  $M_{\rm shared}$ is the mass in common between the HSF structure and
  its counterpart (of mass $M_{\rm initial}$) just prior to merging
  with the main halo.  {\em Top-right panel:} same as top left one,
  but for bound HSF structures.  {\em Bottom-left and bottom-right
    panels:} same as the top ones, but the mass loss is presented as
  the ratio between total final mass and initial mass, for HSF bound
  structures and SUBFIND (bound) structures, respectively.  On all the
  panels, the symbol size is proportional to $M_{\rm initial}$. In
  addition we plot the median value (in red) and the $\sigma$ errors
  calculated in 10 logarithmic bins (in green), with at least 10
  structures per bin.}
\label{belong2}
\end{figure*}

\begin{figure*}
\includegraphics[width=17cm,height=17cm]{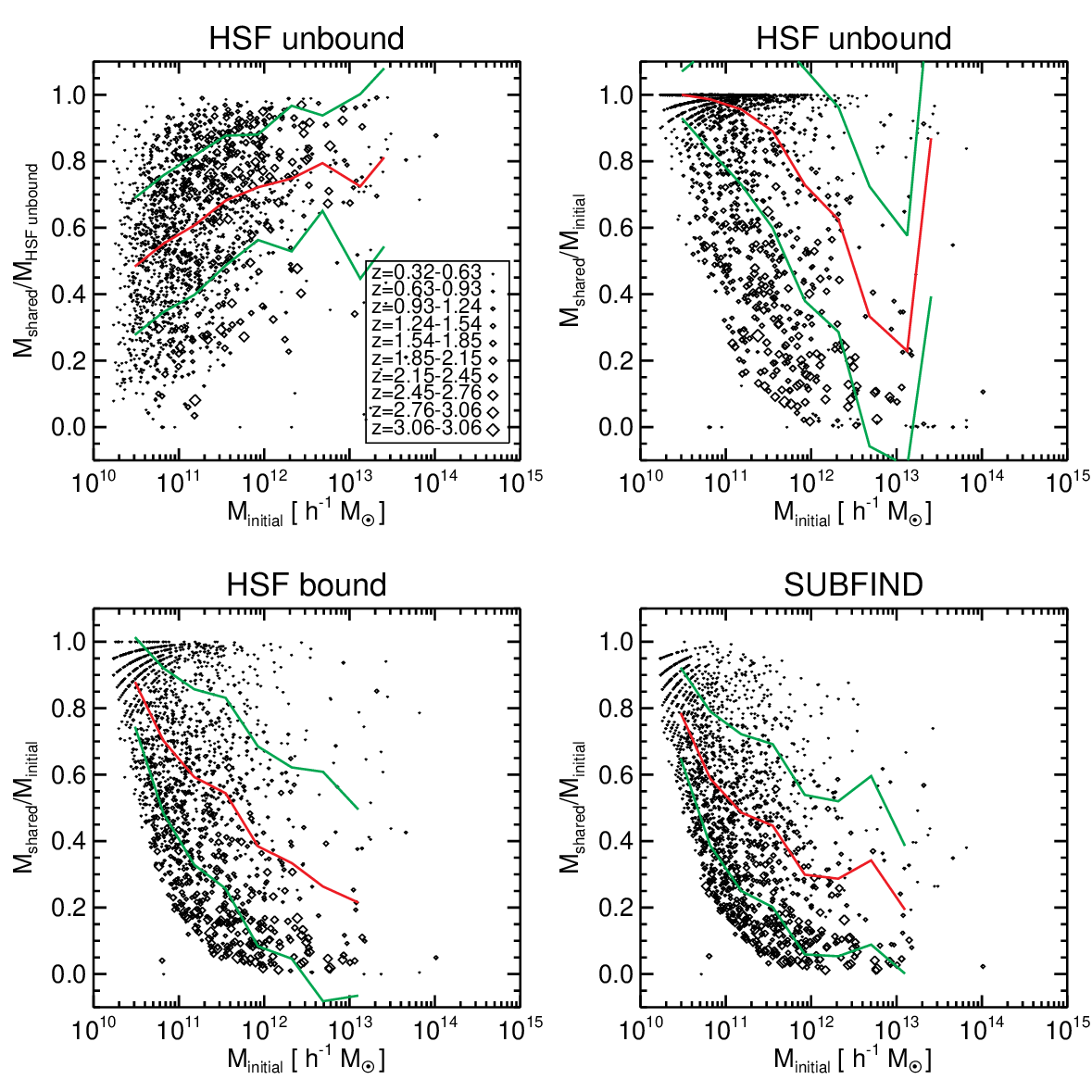}
\caption{ The mass, $M_{\rm shared}$, in common between structures
  detected at present time by HSF and SUBFIND and their counterpart
  --of mass $M_{\rm initial}$-- just prior to merging with the main
  halo, is studied in a fractional way as a function of $M_{\rm
    initial}$.  {\em Top-left:} ratio between $M_{\rm shared}$ and the
  mass of the HSF unbound structure.  {\em Top-right, bottom-left and
    bottom-right panel:} ratio between $M_{\rm shared}$ and $M_{\rm
    initial}$, respectively for HSF unbound, HSF bound and SUBFIND
  structures.  The symbol size is proportional to redshift $z$ of
  merging with the main halo. In addition we plot the median values
  (in red) and the $\sigma$ errors, calculated in 10 logarithmic bins
  (in green), with at least 10 structures per bin.}
\label{belong1}
\end{figure*}

\begin{figure*}
\includegraphics[width=8.5cm,height=8.5cm]{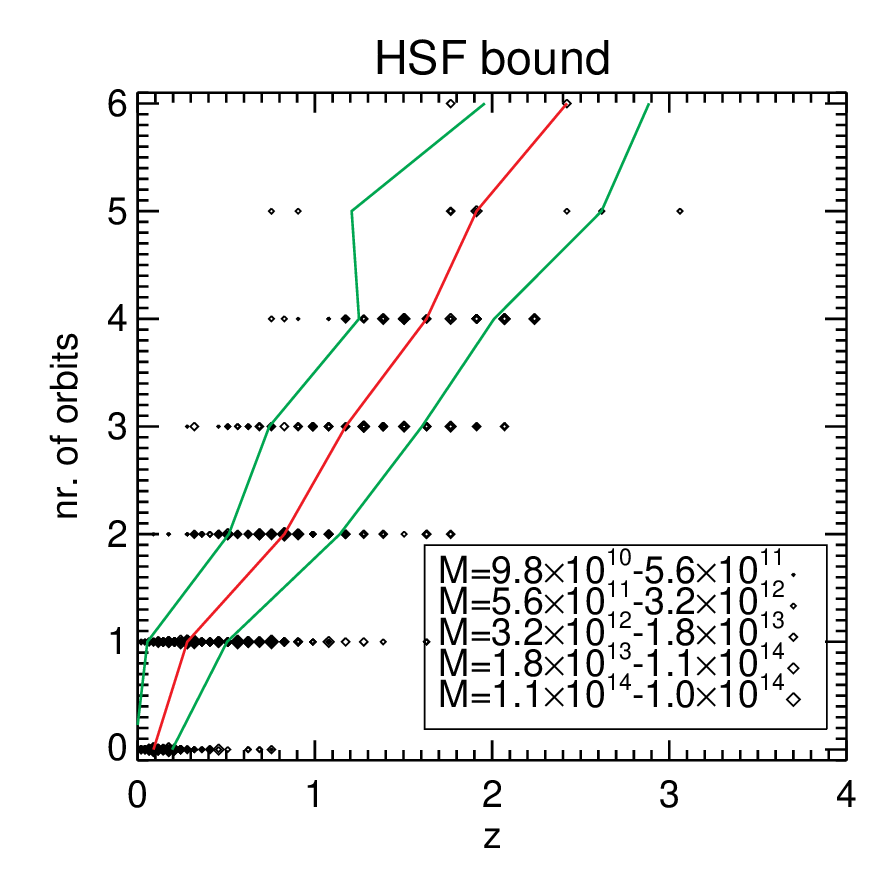}
\caption{Number of orbits each structure made inside the main halo as
  a function of redshift of merging of the structure with the
  halo. The symbol size is proportional to $M_{\rm initial}$,
  expressed in units of $M_{\sun} h^{-1}$.  In addition, we plot the
  median value (in red) and the $\sigma$ errors (in green) calculated
  in 10 logarithmic bins with at least 10 structures in each bin (for
  convenience, binning is performed on $y$ axis).  In our sample the
  structures do not made more than $6$ orbits inside the main halo,
  before they disappear. }
\label{orbits}
\end{figure*}

\begin{figure*}
\includegraphics[width=8.5cm,height=17cm]{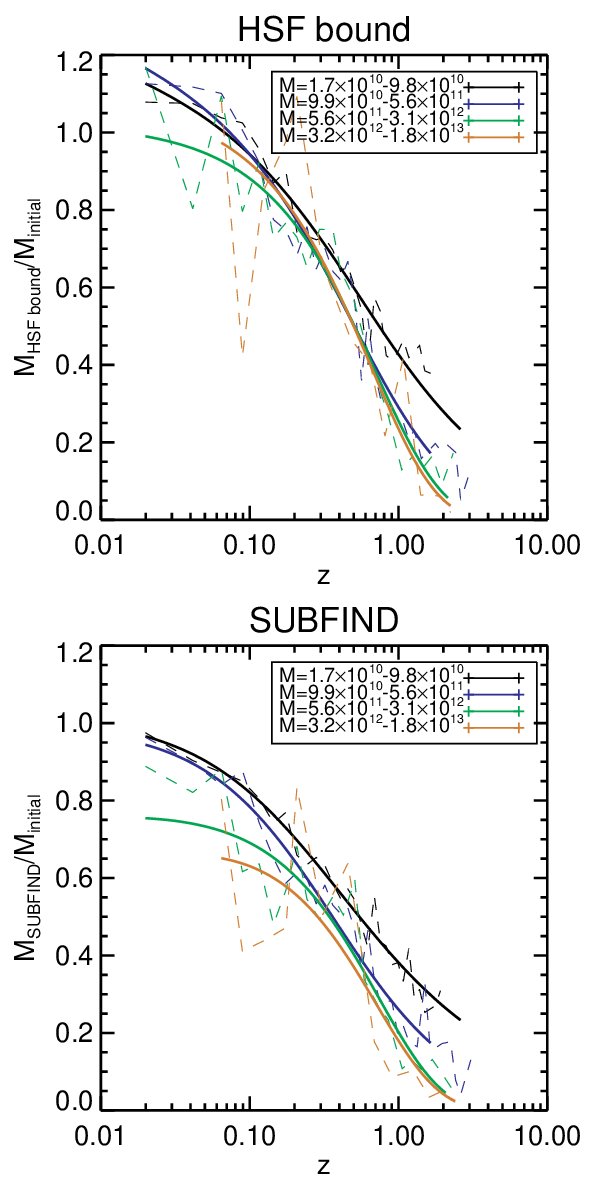}
\caption{Mass loss as a function of redshift of merging $z$ for masses
  binned in 4 logarithmic bins (dotted curves) with its smooth fit
  given by equation (\ref{fit1}) (thick curves).  For each dotted
  curve, the number of bins is equal to $2\sqrt{N}$, where $N$ is the
  number of samples. These two panels are equivalent to bottom panels
  of Fig.~\ref{belong2}.  To make adequate fitting, we perform
  Levenberg-Marquardt least-square minimisation with sigma errors set
  from Poisson noise counting distribution. Masses are expressed in
  units of $M_{\sun} h^{-1}$.}
\label{belong3}
\end{figure*}

To test the performance of HSF quantitatively, one can for example
study the fraction $M_{\rm shared}/M_{\rm initial}$ of particles
inside the initial halo found at the present time in the corresponding
HSF structure, as a function of redshift of merging with the main halo
(top left panel of Figure~\ref{belong2}) or as a function of initial
mass (top right panel of Figure~\ref{belong1}). Indeed, one expects a
strong correlation between the value of $M_{\rm shared}/M_{\rm
  initial}$ and the initial halo mass and the redshift $z$ of merging.
Obviously, the higher $z$, the larger the number of orbits
(Fig.~\ref{orbits}) and the larger the number of particles lost due to
tidal stripping (top left panel of Figure~\ref{belong2}).  At low
redshift, $z \lesssim 0.3$, where the number of orbits is typically
less than one, there is no strong structure evolution in
phase-space and HSF identifies $80-100\%$ of the initial structure
mass (upper left panel of Fig.~\ref{belong2}). There are a few
structures at redshift $z\sim0.1-1.0$ for which HSF can find only a
very small fraction of their initial mass. All of them are tidal
remnants.

The effect of unbinding on the ratio $M_{\rm shared}/M_{\rm initial}$
is shown in the lower left panel of Fig.~\ref{belong1} (as a function
of mass) and upper right panel of Fig.~\ref{belong2} (as a function of
redshift).  Obviously, after unbinding, the fraction of particles
recovered by HSF decreases significantly, even for a small redshift of
merging $z \lesssim 0.3$. Indeed, a significant fraction of the mass
in substructures is contained in tidal tails that are very well
identified by HSF, at least for $z \lesssim 0.3$, but that are not
bound any more to the substructures.  Note that, as expected, the
SUBFIND (bound) substructures are not very different from the HSF
bound ones, except that they contain a slightly smaller fraction of
the mass of the initial structures (lower right panel of
Fig.~\ref{belong1}).

Another important test of our structure finder consists in examining
the fraction of particles inside each HSF structure that is shared
with the initial halo as a function of initial halo mass (top left
panel of Figure~\ref{belong1}).  HSF finds that for massive objects,
$80-90\%$ of the present structure mass belongs to the initial
halo. For smaller structures, the scatter is higher and only around
$40\%$ of the particles found in HSF objects belong to initial halos.
This actually means that for many small structures, the HSF algorithm
attaches more particles than they had before.  This it mainly due to
the fact that we consider for the initial stage only the main part of
the halo and not its substructures: disentangling substructures from
the main halo remains an ambiguous process, and structures identified
at the present time can contain part of the mass of the substructures
inside the initial halo.  In the process, we also do not take into
account particles surrounding the initial halo which were not selected
by the FOF (yet) but were still infalling onto our Millennium test halo
and participate to its background density.  As a result, additional
particles can be associated to the final HSF structure, and this
effect is expected to become stronger if the mass of the identified
structure is small.

Keeping that in mind, we can now study the ratio $M_{\rm X}/M_{\rm
  initial}$ between the total mass of the identified bound structure
and the mass of the initial halo as a function of redshift of merging,
independently of whether the particles are shared between initial and
final stage or not, as shown in the lower panels of Fig.~\ref{belong2}
for HSF (X=HSF bound) and SUBFIND (X=SUBFIND).  The results are not
very different for both codes, except that, as already extensively
argued previously, this ratio is slightly larger for HSF than for
SUBFIND. And because of the effect just discussed above (top left
panel of Fig.~\ref{belong1}), $M_{\rm X}/M_{\rm initial}$ can be
larger than unity at low redshift.

While the correlation between the mass loss due to tidal stripping and
the redshift of merging is quite obvious, the relationship between
mass loss and initial mass is less evident, given the limited amount
of statistical occurrences we analyse here (one single large
cluster-sized Millennium halo).  Figure~\ref{belong1} (right panel and
bottom panels) indicates that the mass loss is significantly smaller
for initially light structures than for initially massive ones. To
demonstrate that unambiguously, we perform a more accurate analysis of
the global mass loss displayed on the lower panels of
Fig.~\ref{belong2}, by fitting analytically the redshift and the mass
dependence. To do so, we divide the initial mass of the structures
into four logarithmic bins. Then for each bin, we fit the mass loss as
a function of redshift in logarithmic coordinates
(Figure~\ref{belong3}) with the following convenient parametric form
\begin{equation}
 \frac{M_{\rm X}}{M_{\rm initial}} = \frac{1}{(z/z_s)^\eta(1+z/z_s)^\gamma}\,.
\label{fit1}
\end{equation}
The best fitting parameters, found by a standard least-square method,
are listed in Table~\ref{fittableHSF}. The fact that mass loss is more
pronounced for more massive objects is clear, and was also to be
expected. This behaviour can simply be explained as follows: small
structures are more strongly bound, because they are more concentrated
\citep[e.g.][]{Angulo2008}, so they do not lose as much mass as large
structures from tidal stripping. Indeed, the largest structures are
less compact and are more sensitive to dynamical friction. As a
result, they are strongly disrupted while they are orbiting around the
main halo.  They also tend to have more radial orbits.

\begin{table*}
\caption{Parameters used in Eq.~(\ref{fit1}) to fit results presented
  in Figure~\ref{belong3}.}
\begin{tabular}{|c|c|c|c|c|c|c|c|}\hline
mass min & mass max & HSF $x_s$ & HSF $\eta$ & HSF $\gamma$ & SUBFIND
$x_s$ & SUBFIND $\eta$ & SUBFIND $\gamma$\\ \hline\hline
$1.7\times10^{10}$ & $9.8\times10^{10}$ & 0.54 & 0.04 & 0.78 & 0.30 & -0.00 & 0.66\\ 
$9.9\times10^{10}$ & $5.6\times10^{11}$ & 1.02 & 0.05 & 1.82 & 0.40 & 0.00 & 1.06\\ 
$5.6\times10^{11}$ & $3.1\times10^{12}$ & 20.11 & 0.00 & 28.31 & 1.46 & -0.03 & 3.01\\ 
$3.2\times10^{12}$ & $1.8\times10^{13}$ & 19188.89 & 0.01 & 28851.01 & 18.93 & -0.07 & 25.61\\ \hline

\label{fittableHSF}
\end{tabular}
\end{table*}



\subsection{Bound subhaloes, tidal tails and tidal streams}
\label{sec:bound}



\begin{figure*}
\includegraphics[width=16cm,height=16cm]{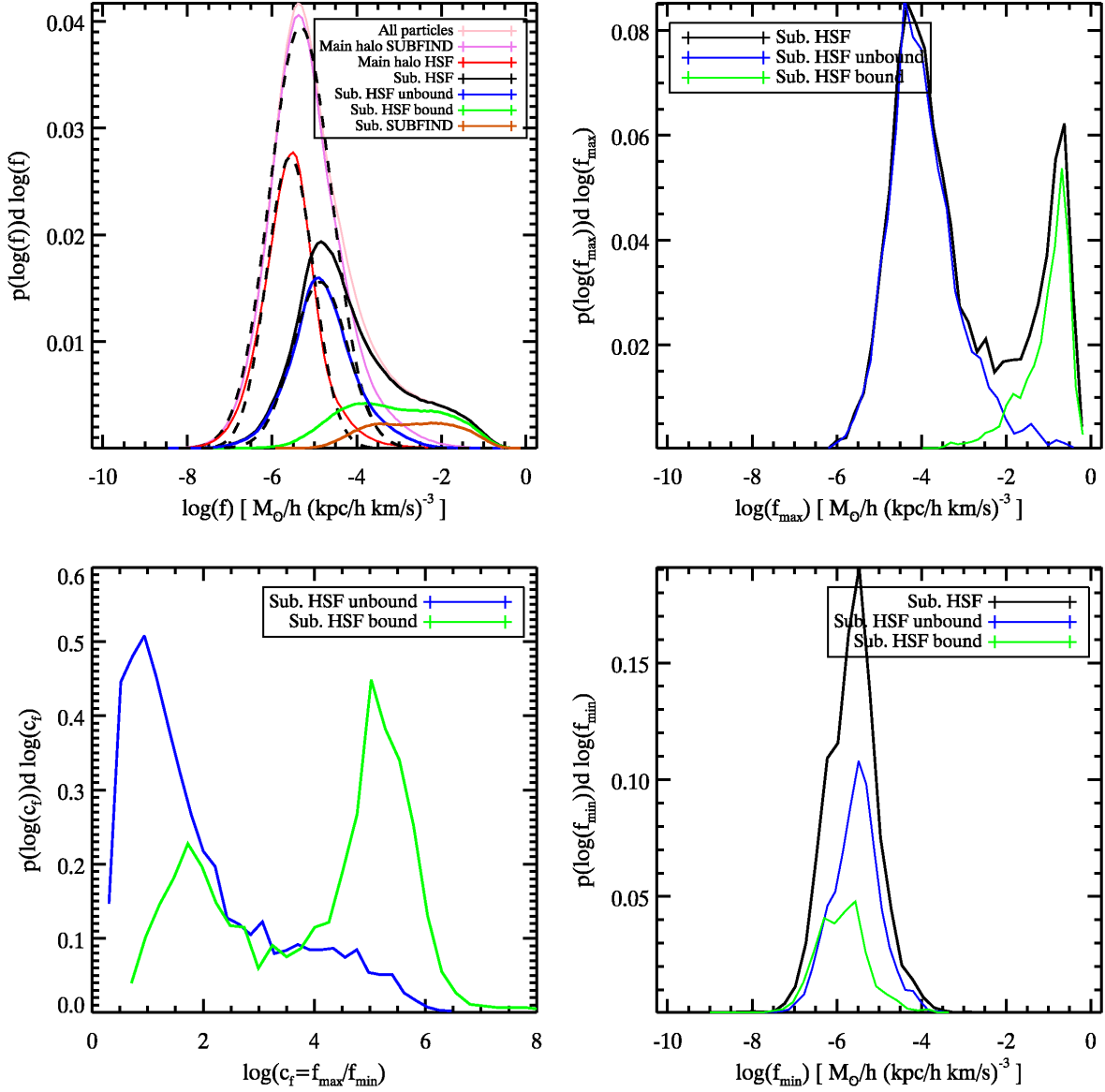}
\caption{Disentangling tidal streams from bound substructures: 
  phase-space density distributions.  {\em Top-left}: distribution function
  of logarithm of phase-space density $f$ estimated for all particles
  belonging to each category of substructures as indicated inside the
  panel (100 logarithmic bins).  The black dashed lines represent the
  best fitted Gaussian functions for the main halo found by SUBFIND,
  the main halo found by HSF and the unbound structures found by HSF
  (including the tidal tails of bound structures).  This means in fact
  that for each of these components, $f$ is lognormal if the fit is
  good. To make adequate fitting, we perform Levenberg-Marquardt
  least-square minimisation with sigma errors set from Poisson Noise
  counting distribution.  {\em Top-right, bottom-right:} distribution
  function of substructures maxima, $f_{\rm max}$, and minima, $f_{\rm
    min}$ (50 logarithmic bins). The substructures are separated into
  unbound components (blue) and bound ones (green), while the black
  curve corresponds to all the substructures.  {\em Bottom-left:}
  distribution function of substructures phase-space ``peakness'',
  defined as $c_f=f_{\rm max}/f_{\rm min}$.}
\label{peak_dist}
\end{figure*}

\begin{figure*}
\includegraphics[width=16cm,height=16cm]{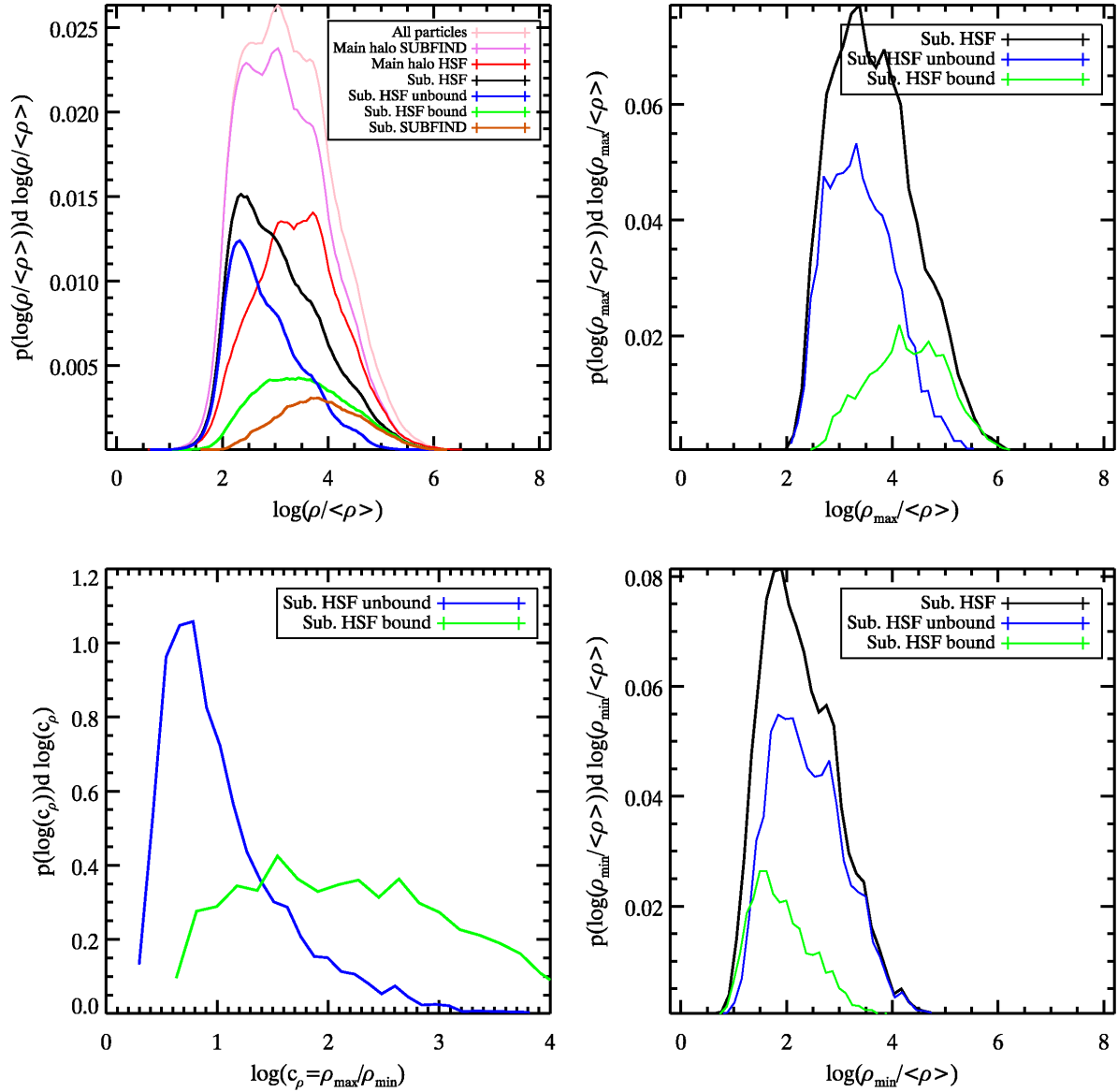}
\caption{Disentangling tidal streams from bound substructures:
  projected 3D density distributions. {\em Top-left}: distribution
  function of normalised density $1+\delta={\rho}/{\langle \rho
    \rangle}$ estimated for all particles belonging to each category
  of substructures as indicated on the panel (100 logarithmic bins).
  {\em Top-right, bottom-right:} distribution function of substructure
  maxima $\rho_{\rm max}/\langle \rho \rangle$, minima $\rho_{\rm
    min}/ \langle \rho \rangle$ (50 logarithmic bins); {\em
    bottom-left:} distribution function of substructures density
  ``peakness'' defined as $c_\rho=\rho_{\rm max}/\rho_{\rm min}$.}
\label{peak_dist1}
\end{figure*}

By studying the merger tree history, we could show that the HSF
algorithm is capable of finding both substructures and their tidal
tails. As we noticed, some of the tidal tails are still connected to
their host substructures, while others are recognised as separate
objects. We now study this bimodality more carefully.

To better separate tidal streams from the bound counterpart of
substructures, we now take a small value of the connectivity
parameter, $\alpha=0.01$. In the following analysis, we shall study
the five following populations:
\begin{enumerate}
 \item Bound substructures found by HSF;
 \item Bound substructures found by SUBFIND;
 \item All HSF substructures (before unbinding);
 \item Unbound HSF structures: substructures found 
       by HSF which disappear during the unbinding 
       process, such as tidal streams;
 \item Bound HSF structures: substructures found 
       by HSF (along with their tidal tails) 
       which remain after the unbinding process.
\end{enumerate}
To analyse substructures properties, we estimate, for each of them,
the phase-space density maximum $f_{\rm max}$ and the minimum $f_{\rm
  min}$. These 2 quantities are measured for the full set of particles
belonging to the substructure prior to the unbinding process. To
better measure high phase-space density peaks, we use the SPH-AM EnBiD
method with 32 neighbours.

In the top-right panel of Figure~\ref{peak_dist}, the distribution of
the$f_{\rm max}$ values of is shown. It is clearly bimodal, and this
property is in fact independent of $\alpha$. It is straightforward to
understand the origin of the bimodality. The local phase-space density
can for instance be approximated as follows \citep{Maciejewski2008}:
\begin{equation}
  \fxv   =  \frac{\rho({\bf x})}{(2\pi)^{3/2} \sigma^3({\bf x})}
  \exp\left\{ -\frac{  [{\bf v}-{\bf v}_0({\bf x})]^2}{2
  \sigma^2({\bf x})} \right\}.
 \label{eq:fap}
\end{equation}
Two cases can the be considered. In the centre of a bound
substructure, i.e.~a standalone structure that survived the unbinding
process, the local density $\rho$ is high and the local velocity
dispersion $\sigma({\bf x})$ is low, which gives a high local
phase-space density maximum. On the other hand, when substructures are
disrupted by strong tidal forces, their local density $\rho$ decreases
and their velocity dispersion, $\sigma({\bf x})$, increases, so their
peak phase-space density is lower.

These statements can be directly checked by unbinding the
substructures found by HSF.  On the top-right panel of
Figure~\ref{peak_dist}, the high density maxima peak of the
distribution is dominated by the bound substructures, as
expected. There is a small fraction of bound substructures for which
$f_{\rm max}$ resides in the lower density maxima regime. We checked
that this happens only for the smallest substructures with around $20$
particles, for which Poisson noise fluctuations start to be
significant. The lower peak of the distribution of values of $f_{\rm
  max}$ is mainly occupied by unbound substructures. There are still
some unbound substructures residing in the higher peak. They have less
than $100$ particles and can be considered as ``Poisson clusters''
(even in Poisson noise it is possible to find high density contrasts).

\begin{table*}
\caption{For each substructure we measure its maximum phase-space 
density $f_{\rm max}$, minimal value $f_{\rm min}$, maximum 
3D density $\rho_{\rm max}$, minimum one $\rho_{\rm min}$, its phase-space
density ``peakness'', $c_f = f_{\rm max}/f_{\rm min}$, 
and 3D density ``peakness'', $c_\rho=\rho_{\rm max}/\rho_{\rm min}$.
Phase-space density is quoted in $\funit$, while $\rho$ is expressed
in units of total average density $\langle \rho \rangle$.}
\begin{tabular}{|c|c|c|c|c|c|c|}\hline
Structure class& $f_{\rm max}$   &$f_{\rm min}$     &$c_{\rm f}$    &$\rho_{\rm max}$&$\rho_{\rm min}$&$c_{\rho}$ \\ \hline\hline
HSF unbound    &$4.0\times10^{-5}$&$3.3\times10^{-6}$&$8.7$          &$2.1\times10^3$  &$69.2$          &$6.1$ \\ 
HSF bound      &$0.2$             &$2.7\times10^{-6}$&$1.1\times10^5$&$1.4\times10^4$  &$31.2$          &$35.5$ \\ \hline
\end{tabular}
\label{strmass}
\end{table*}

Note that the high $f_{\rm max}$ distribution peak is very sharp,
corresponding to $f_{\rm max}\simeq0.2 \funit$.  As already noticed in
\cite{Maciejewski2008}, all the bound substructures present
approximately the same value of $f_{\rm max}$ \citep[see
  also][]{Vass2008}.  This property could be simply an upper bound
imposed by numerical resolution or set by the dynamics, or more
likely a combination of both \citep[e.g.,][]{Binney2004,Vass2008}.
The second peak, dominated by tidal streams, is less pronounced,
although still quite well defined, with a maximum at $f_{\rm
  max}\simeq\ten{4.0}{-5}\funit$, a value about $3.7$ orders of
magnitude lower than what is found for bound structures (all the
values are summarised in table~\ref{strmass}). This shows again the
very clear separation between bound structures and tidal streams.

Another way of separating various substructure populations consists of
measuring their ``peakness'', i.e. the parameter $c_f=f_{\rm
  max}/f_{\rm min}$, where $f_{\rm min}$ is the minimum value of the
phase-space distribution function of the HSF structures (prior to
unbinding). The advantage of the peakness parameter is that, as
opposed to $f_{\rm max}$, it does not depend on a specific choice of
units: a structure with a bound component should present a peakness
parameter very large compared to unity, contrary to a pure tidal
stream. The measurement of $c_f$ is however meaningful only if $f_{\rm
  min}$ is well defined. This is {\em a priori} not obvious as one
expects $f_{\rm min}$ to be very sensitive to local fluctuations in
the noise, which indeed affect the local topology strongly. We checked
that in fact $f_{\rm min}$ is a robust statistics, as suggested by the
rather symmetric behaviour of the curves shown in the bottom-right
panel of Figure~\ref{peak_dist}.  The distribution of measured values
of $c_f$ is shown in the bottom-left panel, and presents of course the
same bimodal nature as $f_{\rm max}$. For instance, one finds that
$c_f$ is typically of the order of $10^5$ for bound structures, while
it is only of the order of $10$ for tidal streams.

\begin{table}
\begin{center}
\caption{Mass in each substructure class compared to the total mass in
  our Millennium test halo.}
\begin{tabular}{|c|c|}\hline
Structure class & mass \\ \hline\hline
SUBFIND bound & $12.4\%$ \\ 
HSF bound & $18.8\%$ \\ 
HSF unbound $\alpha=0.2$ & $22.5\%$ \\ 
HSF unbound $\alpha=0.01$ & $31.6\%$ \\ 
HSF $\alpha=0.2$ & $41.2\%$ \\ 
HSF $\alpha=0.01$ & $50.4\%$ \\ \hline
\end{tabular}
\label{strmass1}
\end{center}
\end{table}

Finally, the top-left panel shows the distribution of measured values
of $f$ for each dark matter particle. In this plot, particles left
over after unbinding HSF substructures, i.e. belonging to the tidal
tails of these substructures, are put on the list of unbound
substructures. The high phase-space density region is dominated by
bound substructures, which is consistent with the observations we made
for the $f_{\rm max}$ distribution function. Note that HSF bound
substructures are more extended into lower phase-space density regions
than SUBFIND ones and are more likely to overlap in terms of density
with unbound streams.  There is in total almost $19\%$ of mass in HSF
bound substructures to compare with $12.4\%$ in SUBFIND ones (see
Table~\ref{strmass1}).  This additional mass in HSF bound
substructures comes from particles which were not found with the
saddle point algorithm working in 3D position space on which SUBFIND
is based. This means that the total bound mass of substructures
strongly depends on the cutting criterion applied to the 3D density
field $\rho$, even with the additional unbinding procedure. 

An examination of the top left panel of Figure~\ref{peak_dist}
suggests that it is possible to perform an optimal cut on $f$, with a
value chosen between $3\times10^{-5}$ and $3\times10^{-4}\, \funit$ so
that most particles with phase-space density above this threshold
belong to bound substructures. Such a criterion was used before in the
literature to mark substructures \citep{Stadel2008}.  Tidal streams
and possibly signatures of caustics occupy the middle range of
phase-space densities, with $31.6\%$ ($22.5\%$ for $\alpha=0.2$) of
the total FOF halo mass belonging to them, which is more than for
bound substructures.  Similarly as for bound substructures, we can set
some lower limit around $10^{-5}$ on the phase-space density and claim
that most particles with higher value of $f$ than this limit belong to
substructures of some kind (bound or unbound). The low phase-space
density regime is indeed dominated by the main halo component.

\begin{table}
\begin{center}
\caption{Best parameters of the Gaussians fitted to the distribution
  function of the logarithm of phase-space density estimated for all
  particles belonging to each category of substructures (top-left
  panel of Figure~\ref{peak_dist}).}
\begin{tabular}{|c|c|c|c|}\hline
Structure class & mean & $\sigma$ & $\chi^2$ error \\ \hline\hline
SUBFIND main halo & $4.7 \times 10^{-6}$ & 0.73 & 7.39 \\ 
HSF main halo & $2.7 \times 10^{-6}$ & 0.55 & 1.11 \\ 
HSF unbound substructures & $1.4 \times 10^{-5}$ & 0.65 & 0.34 \\ \hline
\end{tabular}
\label{gaussfit}
\end{center}
\end{table}

As a final note on the upper-left panel of Figure~\ref{peak_dist}, we
found that the shape of the distribution function of values $f$
observed for each component has interesting properties: it is very
well fit by a lognormal distribution both for the main halo component
found by SUBFIND and HSF, and the unbound substructures found by
HSF. This complements the findings of \cite{Vass2008}, who performed a
similar analysis but used a more {\em add-hoc} approach to separate
various components contributing to the phase-space distribution
function. The best fitting parameters of a Gaussian on the logarithm
of $f$ are given in Table~\ref{gaussfit}. The interpretation of these
results did not seam straightforward to us, so we decided to leave it
for future work. Certainly, a path to follow is to examine the
arguments developed by \cite{Coles1991} to explain the close to
lognormal behaviour of the projected 3D density, $\rho$, relying on the
continuity equation and the positivity of the density.

In practice, in observations of, e.g., X-rays clusters or
gravitational lensing, the 3D density $\rho$ (or its projection on the
sky) is usually used to model dark matter haloes instead of the
phase-space density $f$. To illustrate how the previous results
translate in terms of $\rho$, Figure~\ref{peak_dist1} is similar to
Figure~\ref{peak_dist}, but the calculations are performed for the
normalised density $1+\delta = \rho/\langle \rho \rangle$ instead of
$f$. The 3D density is measured using EnBiD's SPH kernel with $32$
neighbours. Contrary to Figure~\ref{peak_dist}, the distribution
function of values of $\rho_{\rm max}$ (upper right panel) shows only
one peak. The difference between bound and unbound structures shows
much less contrast (see Table~\ref{strmass} for numerical estimates of
typical values of $\rho_{\rm max}$, $\rho_{\rm min}$ and
$c_{\rho}\equiv \rho_{\rm max}/\rho_{\rm min}$).  In particular, bound
structures present a large spread on their 3D local density maxima of
about $2$ orders of magnitudes, in contrast with what happens with
$f_{\rm max}$, and they are more difficult to disentangle from their
unbound counterpart, even with the peakness, $c_{\rho}$, although this
latter quantity seems to have a better separating power than
$\rho_{\rm max}$ (lower left panel).

Interestingly, the particle density distribution diagram
(top-left panel of Figure~\ref{peak_dist1}) is populated in a
different way from what happens in phase-space. In particular, tidal
streams occupy the low density region although they still spread over
a large dynamic range, while the main halo dominates the high density
regime. Bound substructures are rather subdominant and spread over the
whole dynamic range.

\begin{figure*}
\includegraphics[width=16cm,height=21cm]{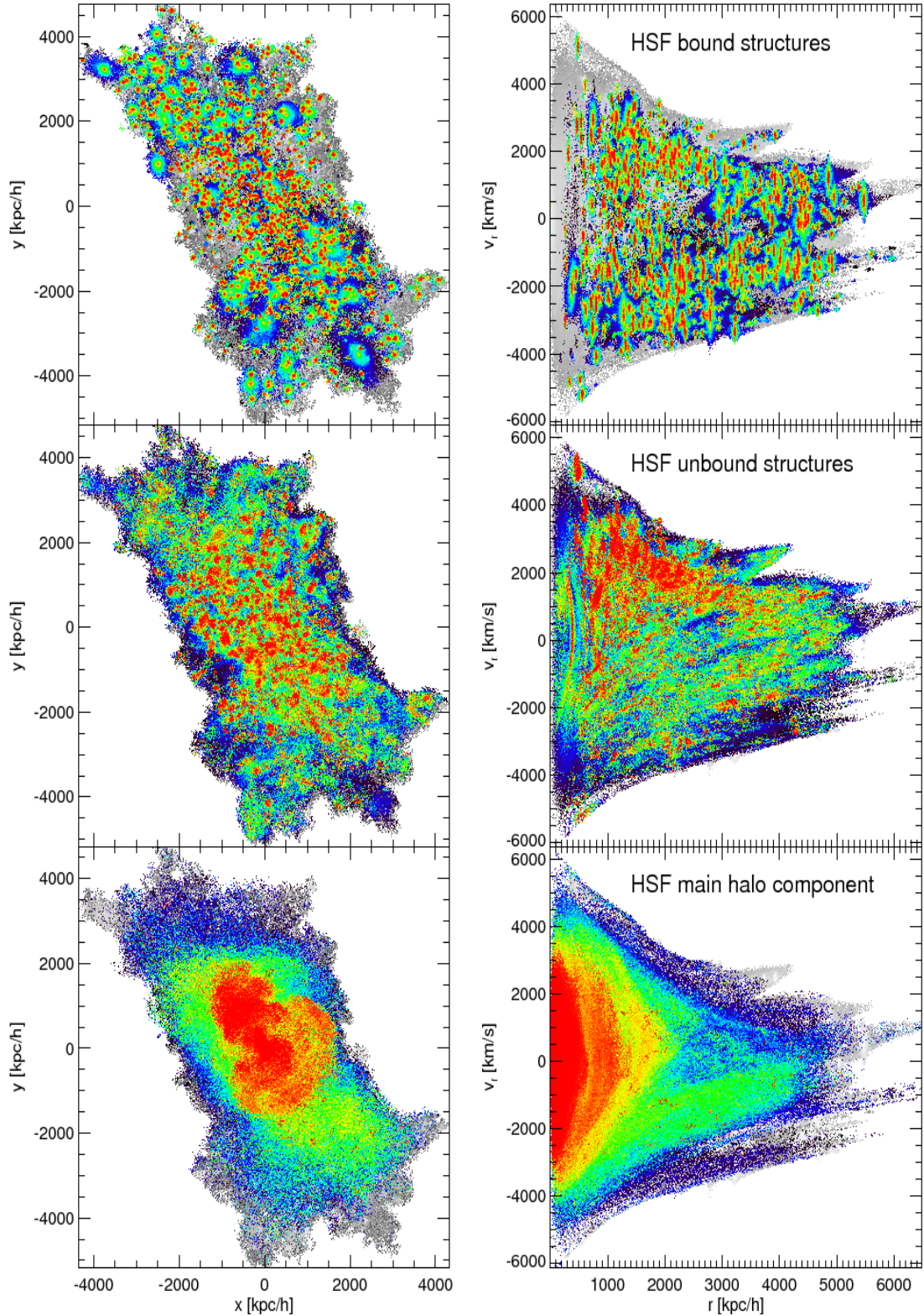}
\caption{Appearance of bound and unbound structures in our Millennium
  test halo.  {\em Top panels:} particles belonging to HSF bound
  structures (so unbound particles are removed).  {\em Middle panels:}
  particles which belong to HSF unbound substructures or are left over
  in the tails of HSF bound substructures after the unbinding process.
  {\em Bottom panels:} the main halo after removal of all
  substructures.  From left to right: $x$--$y$ position space, radius
  $r$--radial velocity $v_r$ phase-space.  The pictures are computed
  in 3 steps as follows: (i) division of space into a
  three-dimensional equally spaced grid with $N=400$ divisions across
  each $x,y,z$ axes, (ii) calculation of the mean density $f$ of all
  particles inside each cell and (iii) projection of this density on
  the $x-y$ plane by taking in each $z$ column the cell with the
  highest density.  To enhance the contrasts, equalisation of the
  histograms in $\log f$ was implemented. }
\label{halo_subs}
\end{figure*}

\begin{figure*}
\includegraphics[width=16cm,height=14cm]{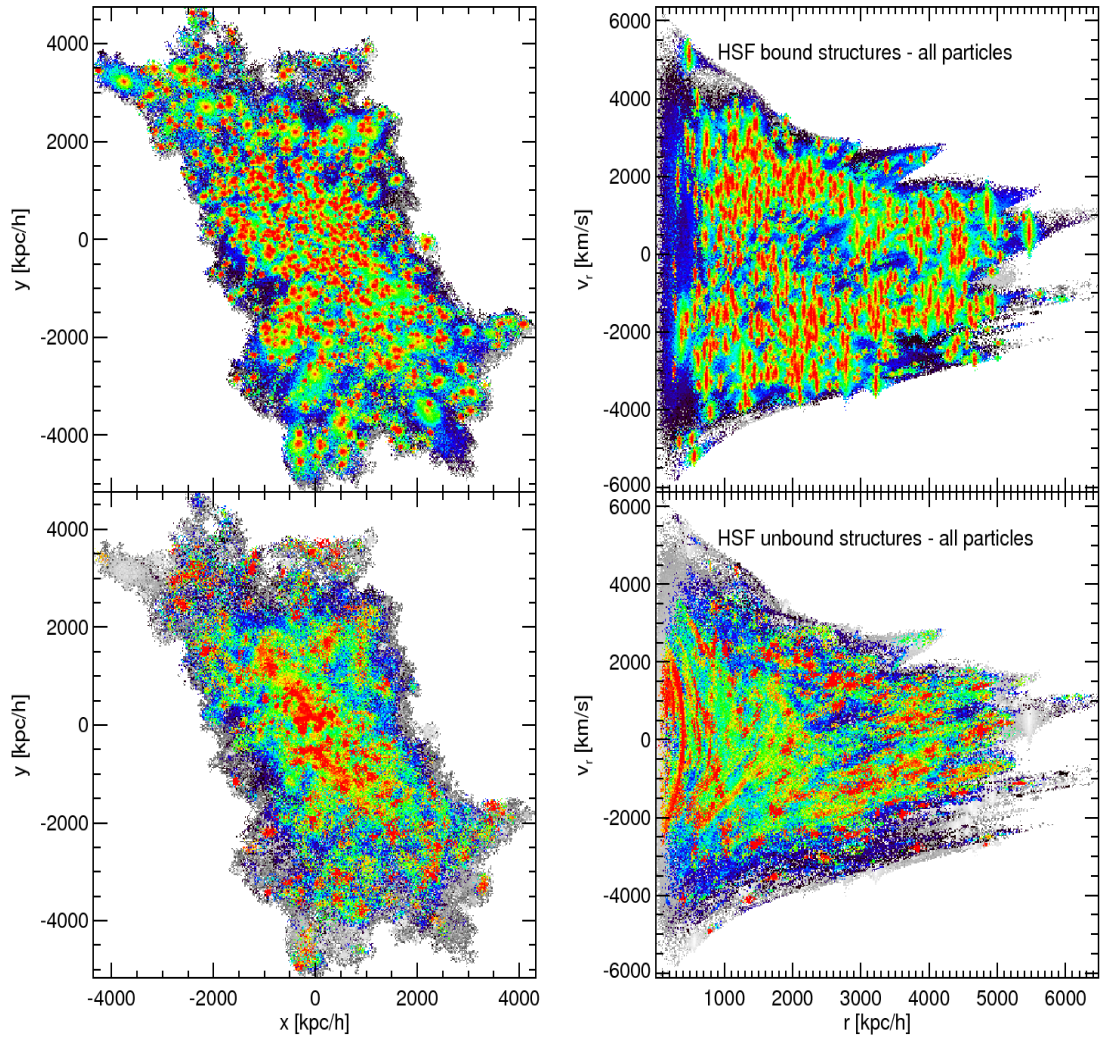}
 \caption{Same as Figure~\ref{halo_subs}, but the set up is slightly
   different: in the {\em top panels}, the HSF structures which are
   seeds of bound structures are displayed entirely, including their
   tidal tails. In the {\em bottom panels,} only HSF structures which
   are unbound are displayed.}
\label{halo_subs2}
\end{figure*}

To complete this section, Figure~\ref{halo_subs} and
Figure~\ref{halo_subs2} show the appearance of bound structures,
unbound ones, and of the smooth part of the halo after removal of all
HSF structures. There is a subtle but significant difference between
the two figures. In Figure~\ref{halo_subs}, the top panels show only
the bound part of the bound substructures, while the top panels of
Figure~\ref{halo_subs2} show the bound structures along with their
tidal tails. In the middle panels of Figure~\ref{halo_subs}, particles
both belonging to unbound structures and particles removed from the
bound structures during the unbinding process are shown. In contrast,
the bottom panels of Figure~\ref{halo_subs2} show only particles
belonging to unbound structures.  This results in an asymmetry in
middle right panel of Figure~\ref{halo_subs}, which reflects the fact
that structures passed through (or nearby) the centre of the halo one
more time in the upper part of the phase-space diagram than in the
lower part. Tidal disruption is indeed more significant and thus
removes particles with higher value of $f$ in the upper part of the
phase-space diagram than in the lower part. Not surprisingly, the
asymmetry disappears in the bottom right panel of
Figure~\ref{halo_subs2}.  Note that bound structures are absent in the
region close to the main halo centre, as expected. Note as well the
rather elongated tidal streams, in particular close to the halo
centre, in the bottom right panel of Figure~\ref{halo_subs2} and the
middle right panel of Figure~\ref{halo_subs}. These are the left overs
of structures disrupted by strong tidal forces. In the bottom right
panel of Figure~\ref{halo_subs}, the main halo still presents, after
cleaning, some filamentary structures, which are parts of tidal tails,
or less likely, signatures of caustics. It can be cleaned even more by
using a smaller value of the connectivity parameter $\alpha$.

\section{Discussions and conclusions}
\label{sec:discussion}

We introduced a new universal multi-dimensional hierarchical structure
finder (HSF) which was employed here to study dark matter structures
in six dimensional phase-space.  The algorithm used, for each
particle, the phase-space density and local neighbourhood estimated
with the SPH method with adaptive metric implemented in the EnBiD
package \citep{Sharma2005}. To detect structures, HSF builds on the
SUBFIND and ADAPTAHOP algorithms, with the introduction of a new
simple but robust cut or grow criterion depending on a single {\rm
  connectivity parameter} $\alpha$.

The main steps of the algorithm are as follows: (i) local phase-space
density maxima are detected and structures around them are grown by
following local gradients up to saddle points; (ii) at each saddle
point level, the density $f$ of each structure is compared to Poisson
noise: the structure is kept if $f$ is $\beta$ times more significant
than the Poisson r.m.s. noise level. At the same time its mass and the
mass of its partner (connected to it through the saddle point) are
measured and the cut or grow criterion is applied: if one structure is
$\alpha$ times smaller than its neighbour, then all particles below
the saddle point are attached to the neighbour. When the two
structures have comparable mass within a factor $\alpha$, they are
both set to grow down as before.  This criterion allows us to better
trace substructures in phase-space, with a good control of the effect
of Poisson noise, which is very important in this rather sparsely
sampled space.

We demonstrated the potential of HSF on a large FOF dark matter halo
taken from the Millennium Simulation. Our tests show that $\beta$ and
especially $\alpha$ are important control parameters. To better study
the smallest possible structures, $\beta$ should be set close to
$0$. The smaller $\alpha$, the more subtle the structures found by the
algorithm. In our analysis, we give preference to $\alpha=0.2$, which
provides a good balance between finding the finest possible
substructures and not overgrowing them. This value of $\alpha$ is
particularly appropriate when an additional binding step is performed.
In contrast, an analysis of tidal tails is best carried out with small
$\alpha$, around $0.01-0.001$, which separates structures into smaller
pieces. It is possible to use it in combination with $\beta=4-10$,
which tends to reconnect the structures together in a consistent way,
to reconstruct tidal tails rather well. A more advanced method of
reconnecting phase-space structures, by using the topology of the
hierarchical tree created by the HSF algorithm is under investigation.

We used the Millennium Simulation merger tree \citep{Springel2005b} to
compare the HSF phase-space structures found at the present time with
the same structures traced back to the time just before they enter the
main halo.  While the best three dimensional algorithms used
presently, such as SUBFIND, manage to find only the main part of bound
structures, HSF is capable of finding more extended bound components
along with their tidal tails.  There is much more information about
structure evolution still stored in phase-space than in 3D and this
information can be potentially fully recovered from the data by a six
dimensional algorithm such as HSF.

The main results of our analysis in time and space domain are the
following:
\begin{itemize}
 
\item
HSF structures contain on average $80-100\%$ of the mass inside the
initial structures up to a redshift of merging $z=0.3-0.4$. This value
drops down to $50\%$ for $z=1$. On the other hand, bound HSF
structures contain on average $80-100\%$ of the mass inside the
initial halos only up to $z=0.09$ and $50\%$ up to $z=0.6$. This shift
in the mass loss is caused by the existence of tidal tails, which are
joined to HSF structures, but do not belong to their bound part. In
other words we can say that HSF is able to reconstruct in most cases
the full dynamical structures which enter the halo at redshifts as
high as $z=0.3-0.4$.

\item
The distribution function of the phase-space density maxima $f_{\rm
  max}$ of HSF structures clearly shows a bimodality. We can explain
it by partitioning the structures into two distinct groups. In the
first group, corresponding to the high phase-space density peak
regime, $f_{\rm max}\approx0.2 \funit$, with a small spread around
that value, there are mostly bound structures. In the second group,
corresponding to a $3$ orders of magnitude smaller phase-space density
regime, $f_{\rm max}\approx3.3\times10^{-5} \funit$, and a larger
spread around this value, there are all unbound structures i.e.~tidal
tails, streams and possibly some caustics. In terms of ``peakness'',
$c_f=f_{\rm max}/f_{\rm min}$, where $f_{\rm min}$ is minimum value of
$f$ in each substructure, this translates into $c_f=1.1\times10^5,8.7$
for bound and unbound structures, respectively.

\item
We noticed, similarly as \cite{Vass2008}, that the distribution
function of the values of $f$ around each dark matter particle is
close to lognormal for the smooth component of the halo and the
unbound part of the substructures (tidal streams).

\item
We found that there is more mass in bound HSF substructures than in
SUBFIND ones. Figure~\ref{mass_profile} shows the cumulative mass of
substructures divided by the total halo mass as a function of
substructure mass.  Around $18.5\%$ of halo mass is stored in bound
HSF structures, and this quantity almost does not depend on $\alpha$.
In comparison, about $12.4\%$ of the mass is stored in SUBFIND bound
structures. The additional mass in HSF bound structures comes mainly
from the fact that subhaloes are better defined in phase-space and are
more extended. However, the set of identified bound substructures is
nearly identical in both methods, and hence the cumulative abundance
of substructures as a function of maximum circular velocity is the
same as well.

\begin{figure}
\includegraphics[width=8.5cm,height=8.5cm]{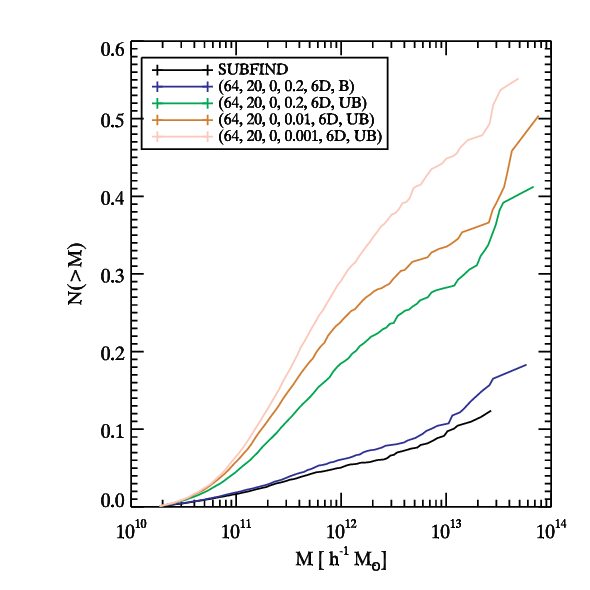}
\caption{Cumulative mass in substructures found by SUBFIND, bound HSF
  method, and HSF with different choices of the connectivity parameter
  $\alpha$.}
\label{mass_profile}
\end{figure}

We note that in our test halo, $41.2\%$ of the mass belongs to
substructures for $\alpha=0.2$, $50.4\% $ for $\alpha=0.01$, and
$55.2\%$ for $\alpha=0.001$. When we substract from these numbers the
contribution of bound structures, we find that $22.9\%-36.4\%$ of halo
mass is stored in unbound structures.  This should be taken into
account when analytical models of halos with substructures are
proposed.

\item
While we would need a larger statistical sample of halos to perform
robust measurements, we noticed that at fixed redshift of merging with
the main halo, small structures tend to lose less mass than larger
ones, in agreement with expectations based on the higher concentration
of smaller halos. Also, we found a strong correlation between mass
loss and the number of orbits a substructure can make inside the main
halo.
\end{itemize}

When we observe our own Galaxy, we do not have access to different
``snapshots'' anymore, in stark difference with the world of
simulations. Instead, we have to be content with the data at the
present time. However, because we now know that our phase-space
structure finder can identify dynamical structures that were bound
before tidal disruption, it can provide totally new insights about the
past dynamical history of our Galaxy. Within the hierarchical
framework, we expect that our Galaxy should be made through the
merging of more than about 100 smaller subcomponents. Comparing
structures in observational data and simulations can be one of the
best tests for the theory of hierarchical galaxy formation, and
provide important constraints on cosmological models such as
$\Lambda$CDM.

Up to now, we studied only the evolution of dark matter, but we can
also similarly study the evolution of baryons in gas and
stars. Galaxies are observed in many different ways, ranging from star
distributions, velocity and chemical properties, to HI measurements,
etc. Our phase-space structure finder with local metric fitting is in
fact implemented in such a way that it can be used in any number of
dimensions, where each dimension can have completely different
physical units. So it is in principle straightforward to use it for
studying galaxy structure evolution in multi-dimensional space with
the appropriate probabilistic weightings to take into account the
noise and holes (missing measurements) in the data hypercube. We think
that such an approach can yield a deeper understanding of galaxy
evolution, and looks especially promising in light of the upcoming
GAIA mission \citep{Gilmore1998} which plans to map the positions of
around one billion stars in our Galaxy.

\section*{Acknowledgments}
We thank S.D.M. White for many important suggestions and
G. Lavaux, R. Mohayaee and D. Weinberg for useful discussions.
Part of this work was completed during a visit of M.M. and 
S.C. at MPA/Garching. This work was completed within the 
framework of the HORIZON project ({\tt www.projet-horizon.fr}). 
M. Maciejewski was funded by the European program MEST-CT-2004-504604
and supported by French ANR (OTARIE).


\begin{thebibliography}{50}
\bibitem[Angulo et al., 2008]{Angulo2008} Angulo R. E., Lacey C. G., Baugh C. M., Frenk C. S., 2008, arXiv:0810.2177
\bibitem[Alard \& Colombi, 2005]{Alard2005} Alard C., Colombi, S., 2005, MNRAS, 359, 123
\bibitem[Aubert,~Pichon \& Colombi, 2004]{Aubert2004} Aubert D., Pichon C., Colombi S., 2004, MNRAS, 352, 376A   
\bibitem[Bertschinger, 1985]{Bertschinger} Bertschinger E., 1985, ApJS, 58, 39
\bibitem[Binney, 2004]{Binney2004} Binney J., 2004, MNRAS, 350, 939B
\bibitem[Cole \& Lacey, 1996]{Cole1996} Cole S., Lacey C., 1996, MNRAS, 281, 716  
\bibitem[Coles \& Jones, 1991]{Coles1991} Coles P., Jones B., 1991, MNRAS, 248, 1
\bibitem[Colombi \& Touma, 2007]{Colombi2007} Colombi S., Touma J., 2007, CNSNS, 03, 012
\bibitem[Davis et al., 1985]{Davis1985} Davis M., Efstathiou G., Frenk C. S., White S. D. M., 1985, ApJ, 292, 371 
\bibitem[Diemand et al., 2006]{Diemand2006} Diemand J., Kuhlen M., Madau P., 2006, ApJ, 649, 1 
\bibitem[Eisenstein \& Hut, 1998]{Eisenstein1998} Eisenstein D. J., Hut P., 1998, ApJ, 498, 137 
\bibitem[Gelb \& Bertschinger, 1994]{Gelb1994} Gelb J., Bertschinger E., 1994, ApJ, 436, 467 
\bibitem[Gilmore et al., 1998]{Gilmore1998} Gilmore G. F., Perryman M. A., Lindegren L., Favata F., Hoeg E., et al., 1998, SPIE, 3350, 541G
\bibitem[Governato et al., 1997]{Governato1997} Governato F., Moore B., Cen R., Stadel J., Lake G., Quinn T., 1997, NewA, 2, 91 
\bibitem[Hinshaw et al., 2008]{WMAP5} Hinshaw G., Weiland J. L., Hill R. S., Odegard N., Larson D., Bennett C. L. et al., 2008,	arXiv:0803.0732
\bibitem[Kim \& Park, 2006]{Kim2006} Kim J. \& Park, C., 2006, ApJ, 639, 600 
\bibitem[Klypin et al., 1999]{Klypin1999} Klypin A., Gottlober S., Kravtsov A. V., Khokhlov, A. M., 1999, ApJ, 516, 530 
\bibitem[Lacey \& Cole, 1994]{Lacey1994} Lacey C., Cole S., 1994, MNRAS, 271, 676 
\bibitem[Luki\'{c} et al., 2008]{Lukic2008} Luki\'{c} Z., Reed D., Habib S., Heitmann K., 2008, arXiv:0803.3624 
\bibitem[Maciejewski et al., 2008]{Maciejewski2008} Maciejewski M., Colombi S., Alard C., Bouchet F., Pichon C., 2008, MNRAS, in press (arXiv:0810.0504)
\bibitem[Massey et al., 2007]{COSMOS} Massey R., Rhodes J., Ellis R., Scoville N., Leauthaud A. et al., 2007, Nature, 445, 286M
\bibitem[Mohayaee \& Salati, 2008]{Roya2008} Mohayaee R., Salati P., 2008, arXiv:0801.3271
\bibitem[Natarajan,~De Lucia \& Springel, 2007]{Natarajan2007} Natarajan P., De Lucia G., Springel V., 2007, MNRAS, 376, 180
\bibitem[Navarro,~Frenk \& White, 1997]{NFW} Navarro J. F., Frenk C. S., White S. D. M., 1997, ApJ, 490, 493
\bibitem[Neyrinck et al., 2005]{Neyrinck2005} Neyrinck M. C., Gnedin N. Y., Hamilton A. J. S., 2005, MNRAS, 356, 1222 
\bibitem[Rubin \& Ford, 1970]{Rubin1970} Rubin V. C., Ford. W. K. J., 1970, ApJ, 159, 379
\bibitem[Sharma and Steinmetz, 2005]{Sharma2005} Sharma, S.,  Steinmetz, M., 2006, MNRAS 373, 1293 
\bibitem[Springel et al., 2001]{Springel2001} Springel V., White S. D. M., Tormen G., Kauffmann G., 2001, MNRAS, 328, 726 
\bibitem[Springel et al., 2005a]{Springel2005} Springel V., White S. D. M., Jenkins A., Frenk, C. S., et al.,  2005a, Nature, 435, 629
\bibitem[Springel et al., 2005b]{Springel2005b} Springel V., White S. D. M., Jenkins A., Frenk, C. S., et al., 2005b, Nature, 435, 629, 
             Supplementary Information (astro-ph/0504097v2)
\bibitem[Springel, 2005]{Springel2005GADGET2} Springel V., 2005, MNRAS, 364, 1105 
\bibitem[Springel et al., 2008]{Springel2008} Springel V., White S. D. M., Frenk C. S., Navarro J. F., Jenkins, A., et al., 2008, arXiv:0809.0894
\bibitem[Stadel et al., 2008]{Stadel2008} Stadel J., Potter D., Moore B. Diemand, J., Madau P. et al. astro-ph/0808.2981v2 
\bibitem[Van Waerbeke et al., 2000]{lensing} Van Waerbeke, L., Mellier Y., Erben T., Cuillandre J. C., Bernardeau F. et al., 2000, A\&A, 358, 30
\bibitem[Vass et al., 2008]{Vass2008} Vass H.M., Valluri M., Kravtsov A.V., Kazantzidis S., 2008, submitted to MNRAS (arXiv:0810.0277)
\bibitem[Vogelsberger et al., 2008]{Vogelsberger2008} Vogelsberger M., White S.D.M., Helmi A., Springel V., 2008, MNRAS, 385, 236
\bibitem[White \& Rees, 1979]{White1978} White S. D. M., Rees M. J., 1978, MNRAS, 183, 341
\bibitem[White \& Vogelsberger, 2008]{White2008} White S.D.M.,  Vogelsberger M., 2008, arXiv:0809.0497
\bibitem[Zwicky, 1933]{Zwicky1933} Zwicky F., 1933, Helvetica Physica Acta, 6, 110

\end{thebibliography}
\end{document}